\documentclass[fleqn,final,5p,times]{elsarticle}

\usepackage{amssymb}
\usepackage{amsthm}

\usepackage{amsmath}
\usepackage{booktabs}
\usepackage{calc}
\usepackage{natbib}
\usepackage{bm}
\usepackage{siunitx}
\usepackage{xfrac}
\usepackage{nth}

\usepackage{tikz}
\usetikzlibrary{colorbrewer}
\usetikzlibrary{positioning}
\usetikzlibrary{shapes.geometric}
\usetikzlibrary{arrows}

\usepackage{pgfplots}
\usepgfplotslibrary{groupplots}
\usepgfplotslibrary{colorbrewer}
\usepgfplotslibrary{fillbetween}

\pgfplotsset{
compat=1.16,
cycle list/Paired,
grid=both,
  grid style={line width=.1pt, draw=gray!5},
  major grid style={line width=.2pt,draw=gray!25},
siunitxlabels/.style={
    /pgfplots/typeset ticklabel/.code={{\pgfmathparse{\tick}$\num[zero-decimal-to-integer]{\pgfmathresult}$}},
  },
}

\newcommand{\cred}{     230,  25,  75}
\newcommand{\cgreen}{    60, 180,  75}

\newcommand{\cblue}{      0, 130, 200}
\newcommand{\corange}{  245, 130,  48}
\newcommand{\ccyan}{     70, 240, 240}
\newcommand{\cmagenta}{ 240,  50, 230}

\newcommand{\cteal}{      0, 128, 128}
\newcommand{\clavender}{220, 190, 255}
\newcommand{\cbrown}{   170, 110,  40}
\newcommand{\cbeige}{   255, 250, 200}
\newcommand{\cmaroon}{  128,   0,   0}

\newcommand{\cnavy}{      0,   0, 128}
\newcommand{\cgrey}{    128, 128, 128}

\definecolor{color0}{RGB}{\cmaroon}
\definecolor{color1}{RGB}{\corange}
\definecolor{color2}{RGB}{\cteal}
\definecolor{color3}{RGB}{\ccyan}
\definecolor{color4}{RGB}{\cgrey}
\definecolor{color5}{RGB}{\cnavy}
\definecolor{color6}{RGB}{\cred}
\definecolor{color7}{RGB}{\cbrown}
\definecolor{color8}{RGB}{\cmagenta}
\definecolor{color9}{RGB}{\cgreen}
\definecolor{color10}{RGB}{\cbeige}
\definecolor{color11}{RGB}{\cblue}
\definecolor{color12}{RGB}{\clavender}

\definecolor{colora}{RGB}{\cblue}
\definecolor{colorb}{RGB}{\cbrown}
\definecolor{colorc}{RGB}{\cgreen}
\definecolor{colord}{RGB}{\cred}
\definecolor{colore}{RGB}{\corange}
\definecolor{colorf}{RGB}{\cteal}

\newcommand{\ppvec}[1]{\bm{#1}}
\newcommand{\code}[1]{\textsc{#1}}
\newcommand{\ppRe}{\ensuremath{\mathrm{Re}}}
\newcommand{\ppMa}{\ensuremath{\mathrm{Ma}}}

\graphicspath{{fig/}}

\usepackage{xcolor}

\usepackage[hidelinks]{hyperref}

\usepackage{tabularx}
\usepackage{subcaption}
\usepackage[capitalise]{cleveref}
\usepackage{xspace} 

\newcommand{\galexi}{GA\-L{\AE}\-XI\xspace}

\newcommand{\numTwoPlaces}[1]{\num[round-mode=places,round-precision=2]{#1}}

\usepackage{algpseudocode}
\usepackage{algorithm}
\algnewcommand\algorithmicswitch{\textbf{switch}}
\algnewcommand\algorithmiccase{\textbf{case}}
\algnewcommand\algorithmickernel{\textbf{kernel}}
\algdef{SE}[KERNEL]{Kernel}{EndKernel}[2]{\algorithmickernel\ \code{#1}\ (#2)}{\algorithmicend\ \algorithmickernel}\algdef{SE}[SWITCH]{Switch}{EndSwitch}[1]{\algorithmicswitch\ (#1)}{\algorithmicend\ \algorithmicswitch}\algdef{SE}[CASE]{Case}{EndCase}[1]{\algorithmiccase\ #1}{\algorithmicend\ \algorithmiccase}\algtext*{EndSwitch}\algtext*{EndCase}\algrenewcommand\algorithmicindent{1.0em}

\journal{}

\begin{document}

\begin{frontmatter}

\title{\galexi: Solving complex compressible flows with high-order discontinuous Galerkin methods on accelerator-based systems}

\author[label1,label2]{Marius Kurz\fnref{fn1}}
\ead{marius.kurz@iag.uni-stuttgart.de}

\author[label1]{Daniel Kempf\fnref{fn1}\corref{cor1}}
\ead{daniel.kempf@iag.uni-stuttgart.de}

\author[label1]{Marcel Blind}
\ead{marcel.blind@iag.uni-stuttgart.de}

\author[label1]{Patrick Kopper}
\ead{patrick.kopper@iag.uni-stuttgart.de}

\author[label3]{Philipp Offenh{\"a}user}
\ead{philipp.offenhaeuser@hpe.com}

\author[label1]{Anna Schwarz}
\ead{anna.schwarz@iag.uni-stuttgart.de}

\author[label1]{Spencer Starr}
\ead{spencer.starr@iag.uni-stuttgart.de}

\author[label1]{Jens Keim}
\ead{jens.keim@iag.uni-stuttgart.de}

\author[label1]{Andrea Beck}
\ead{andrea.beck@iag.uni-stuttgart.de}

\fntext[fn1]{M. Kurz and D. Kempf contributed equally and cordially agree to share first authorship.}
\address[label1]{Institute of Aerodynamics and Gas Dynamics, University of Stuttgart, Pfaffenwaldring 21, 70569 Stuttgart, Germany}
\address[label2]{Centrum Wiskunde \& Informatica (CWI), Science Park 123, 1098 XG Amsterdam, The Netherlands}
\address[label3]{Hewlett Packard Enterprise (HPE), Herrenberger Straße 140, 71034 Böblingen, Germany}

\cortext[cor1]{Corresponding author}

\begin{abstract}

  This work presents \galexi as a novel, energy-efficient flow solver for the simulation of compressible flows on unstructured hexahedral meshes leveraging the parallel computing power of modern Graphics Processing Units (GPUs).
\galexi implements the high-order Discontinuous Galerkin Spectral Element Method (DGSEM) using shock capturing with a finite-volume subcell approach to ensure the stability of the high-order scheme near shocks.
This work provides details on the general code design, the parallelization strategy, and the implementation approach for the compute kernels with a focus on the element local mappings between volume and surface data due to the unstructured mesh.
  The scheme is implemented using a pure distributed memory parallelization based on a domain decomposition, where each GPU handles a distinct region of the computational domain.
  On each GPU, the computations are assigned to different compute streams which allows to antedate the computation of quantities required for communication while performing local computations from other streams to hide the communication latency. This parallelization strategy allows for maximizing the use of available computational resources.
This results in excellent strong scaling properties of \galexi up to 1024 GPUs if each GPU is assigned a minimum of one million degrees of freedom.
To verify its implementation, a convergence study is performed that recovers the theoretical order of convergence of the implemented numerical schemes.
  Moreover, the solver is validated using both the incompressible and compressible formulation of the Taylor--Green-Vortex at a Mach number of 0.1 and 1.25, respectively.
  A mesh convergence study shows that the results converge to the high-fidelity reference solution and that the results match the original CPU implementation.
Finally, \galexi is applied to a large-scale wall-resolved large eddy simulation of a linear cascade of the NASA Rotor 37.
  Here, the supersonic region and shocks at the leading edge are captured accurately and robustly by the implemented shock-capturing approach.
It is demonstrated that \galexi requires less than half of the energy to carry out this simulation in comparison to the reference CPU implementation.
This renders \galexi as a potent tool for accurate and efficient simulations of compressible flows in the realm of exascale computing and the associated new HPC architectures.
\end{abstract}


\end{frontmatter}

\section{Introduction}The computational sciences have become an essential driver for understanding the dynamics of complex, nonlinear systems ranging from the dynamics of earth's climate~\cite{lynch2008origins} to obtaining information about a patient's characteristic blood flow to derive personalized approaches in medical therapy~\cite{gundelwein2018personalized}.
While these successes also rely on significant breakthroughs in the development of numerical methods and physical models, a major portion can be ascribed to the exponential increase in available computing power, which has allowed simulating increasingly large and complex problems over the last decades.
However, the corresponding process of shrinking transistors from generation to generation has become increasingly challenging and the resulting gains in performance have diminished in recent years~\cite{lundstrom2022moore}.
As a consequence, the community is moving towards accelerator chips, which do not serve as a one-size-fits-all hardware like traditional CPUs.
These chips are specialized to yield better performance and efficiency for specific tasks, such as workloads in artificial intelligence, video encoding or cryptography.
This also shows in the field of high-performance computing~(HPC), where nine out of the ten fastest supercomputers listed in the most recent TOP500~\cite{top500} from November 2023 employ some form of accelerator.
In the most recent GREEN500~\cite{green500} list, which focuses on sustainability in terms of energy invested per computation, all of the ten most efficient HPC systems employ GPU accelerators.

However, such accelerators generally differ considerably from general-purpose CPUs in terms of hardware design, as well as the working principle.
As a consequence, using accelerators oftentimes not only requires rewriting and redesigning large portions of existing code to make efficient use of such hardware, but also might change which numerical algorithms are most efficient for a specific task.
This poses significant challenges for legacy HPC codes due to the considerable effort required to migrate the existing codebase to hardware accelerators.
This issue is particularly pervasive in the field of Computational Fluid Dynamics~(CFD), where scale-resolving simulations of turbulent flows generally require significant HPC resources. Here, modern high-order discretization methods such as Discontinuous Galerkin~(DG) and Flux Reconstruction~(FR) schemes have become popular due to their computational efficiency for such multi-scale problems and their excellent scaling properties on HPC systems.

Over the last two decades, these advantages have led to the development of a rich landscape of high-order HPC solvers for scale-resolving CFD simulations.
Here, spectral element discretizations have seen particular interest and have been implemented for instance in Nektar++~\cite{moxey2020nektar} and HORSES3D~\cite{ferrer2023horses}, which can solve both the compressible and incompressible Navier--Stokes equations~(NSE), and the DG solver ExaDG~\cite{arndt2020exadg}, which solves the incompressible NSE.
While these solvers are written in traditional HPC languages like C++ and Fortran, the solver Trixi.jl~\cite{ranocha2022adaptive} implements multiple DG discretizations in Julia to combine the advantages of high computational performance with the flexibility and ease of implementation provided by modern Python-style languages.
In addition to DG-based approaches, also other high-order methods have been widely adopted using for instance weighted essentially non-oscillatory (WENO) finite volume (FV) schemes as implemented in UCNS3D \cite{antoniadis2022ucns3d} or high-order finite difference methods with WENO-type shock capturing utilized in the STREAmS~\cite{bernardini2021streams,bernardini2023streams} solver.

The need to adapt the existing and established code bases to the new GPU-focused HPC architectures has already been considered in the CFD community.
One of the most established high-order codes for incompressible and weakly compressible flow is Nek5000~\cite{fischer2007nek5000}.
The first effort to port Nek5000 to accelerators was reported in 2015~\cite{markidis2015openacc}, where a barebones version of the code was adapted for GPUs using the OpenACC library.
Full GPU support was then offered by its successor NekRS~\cite{fischer2022nekrs}, which is based on the Open Concurrent Computing Abstraction~(OCCA)~\cite{medina2014occa}.
Similarly, Neko~\cite{jansson2024neko} was implemented from scratch using modern, object-orientated Fortran and abstraction layers to support multiple hardware backends.
While the previous codes focus mainly on incompressible and weakly compressible flows, pyFR~\cite{witherden2015heterogeneous} also solves the compressible NSE on unstructured meshes using the FR approach.
Moreover, it is written in Python and relies on code generation to support multiple computing backends including accelerators and provides excellent scaling properties on HPC systems.
Similarly, the deal.II~\cite{arndt2021deal} and MFEM~\cite{anderson2021mfem} libraries provide DG discretizations to solve the compressible NSE and both have added GPU support in recent years.

In this work, we present \galexi\footnote{\url{https://github.com/flexi-framework/galaexi/}} as a GPU-accelerated solver for hyperbolic-parabolic conservation laws with special emphasis on compressible flows.
The numerical simulation of compressible flows is highly relevant for a large number of problems, e.g.,\ from the aviation industry or aeroacoustics.
\galexi builds on the well-established FLEXI solver~\cite{krais2021flexi} and inherits the majority of its features and its extensive pre- and post-processing suite that is designed for large-scale applications.
Hence, \galexi implements multiple flavors of the Discontinuous Galerkin Spectral Element Method (DGSEM) and can handle fully unstructured, hexahedral, curved, high-order meshes to account for complex geometries.
Moreover, multiple stabilization techniques are implemented to ensure the stability of the scheme in underresolved simulations and in the vicinity of shocks using shock capturing schemes based on localized finite-volume~(FV) subcell approaches.
We consider this focus on compressible, transonic flow through the combination of high-order accuracy and provable stability near shocks together with the efficient implementation of this hybrid discretization on GPU systems as one of the unique features of \galexi that sets it apart from existing solvers.
The user interface of \galexi is deliberately kept compatible to FLEXI, such that \galexi can serve as a drop-in replacement to run existing simulation setups on GPU systems without modifications.

This work contributes the following aspects to the challenging but necessary steps for the transition to exascale HPC architectures in CFD.
It provides insights into the suitability of DGSEM for GPU acceleration and quantifies the gains in performance and efficiency that can be expected for explicit, high-order DG methods when moving from traditional CPUs to GPUs.
It also provides practical guidelines on how existing codebases can be ported to GPU hardware and proposes parallelization concepts for achieving parallel efficiency on HPC hardware with high-order schemes.
Furthermore, the savings in the context of energy-to-solution are discussed in particular and can serve as a point of reference in terms of energy efficiency.

This work is organized as follows.
First, \cref{sec:numerics} introduces the governing equations and the numerical methods implemented in \galexi.
Based on this, \cref{sec:implementation} provides details on the parallelization strategy and the implementation of the compute kernels.
The resulting performance and scaling abilities of \galexi are presented and discussed in \cref{sec:scaling}.
The implementation of the numerical scheme is verified in \cref{sec:validation}, demonstrating the theoretical convergence rates of the numerical methods and accurate results for the incompressible and compressible formulations of the Taylor--Green-Vortex~(TGV) test case.
To demonstrate the applicability to applications of relevance, \galexi is employed in \cref{sec:application} to compute a large-scale, wall-resolved LES of the NASA Rotor~37~\cite{Reid1978} test case.
\cref{sec:conclusion} summarizes the major results of the paper and provides an outlook on further developments.

\section{Numerical Methods}\label{sec:numerics}

\galexi is implemented as a general solution framework for hyperbolic-parabolic conservation equations, similar to the FLEXI framework, but exhibits a particular focus on the compressible Navier--Stokes equations (NSE), which are introduced in \cref{sec:governing_equations}.
The DGSEM with its temporal and spatial high-order accurate discretization will be introduced in \cref{sec:dgsem}, followed by the compatible sub-cell shock capturing scheme in \cref{sec:shock_capturing}.

\subsection{Governing Equations}\label{sec:governing_equations}
\galexi is used to solve the compressible NSE, which describe the evolution of the conserved variables \mbox{$\ppvec{U}(\ppvec{x},t) = (\rho,\rho\ppvec{u},\rho e)^T$} which are comprised of the density, momentum, and energy density, respectively, at each position in space $\ppvec{x}$ and time $t$.
The NSE can be derived by enforcing the conservation of mass, momentum, and energy across an infinitesimal control volume.
This yields the evolution equations of the conserved variables in differential form as
\begin{align}
  \frac{\partial \rho}{\partial t} &+ \nabla \cdot \left(\rho \ppvec{u}\right) &&= 0 ,\\
  \frac{\partial \rho\ppvec{u}}{\partial t} &+ \nabla \cdot \left(\rho \ppvec{u}\otimes \ppvec{u} + p \ppvec{I} - \ppvec{\tau}\right) &&= \ppvec{0} ,\\
  \frac{\partial \rho e}{\partial t} &+ \nabla \cdot \left(\ppvec{u}\left(\rho e+p\right)-\ppvec{\tau}\cdot\ppvec{u}+\ppvec{q}\right) &&= 0 ,
  \label{eq:nse}
\end{align}
where $p$ denotes the static pressure, $\ppvec{I}$ the identity matrix, and $\ppvec{0}$ the zero vector.
Assuming a Newtonian fluid and Fourier's law of thermal conduction yields the stress tensor $\ppvec{\tau}$ and heat flux $\ppvec{q}$ as
\begin{align}
  \ppvec{\tau} &= \mu \left( \nabla \ppvec{u} + \nabla \ppvec{u}^T -  \frac{2}{3} \left(\nabla\cdot \ppvec{u}\right)\ppvec{I}\right) ,\\
  \ppvec{q}    &= - \lambda\: \nabla T.
\end{align}
Here, $\mu$ denotes the dynamic viscosity of the fluid and $\lambda$ denotes its heat conductivity.
Both are material properties of the specific fluid and depend in the general case on the fluid's local state.
Hence, both quantities cannot be considered constant in the general case.
In this work, we assume the viscosity to follow Sutherland's law~\cite{sutherland1893lii}, which postulates a dependency of the viscosity on the temperature of the form
\begin{equation}
  \mu(T) = \mu_{ref} \frac{1.4042\,(T/T_{ref})^{3/2}}{T/T_{ref}+0.4042},
  \label{eq:sutherland}
\end{equation}
where $\mu_{ref}$ is the viscosity at the reference temperature $T_{ref}$.
Based on this, the thermal conductivity can be computed as
\begin{equation}
  \lambda=\frac{\gamma R}{\gamma-1}\frac{\mu}{\mathrm{Pr}},
\end{equation}
with $\gamma$ as the ratio of specific heats, $R$ denoting the specific gas constant, and $\mathrm{Pr}$ as the dimensionless Prandtl number, which is assumed in the following to be constant with $\mathrm{Pr}=0.71$.

Lastly, the equation-of-state~(EOS) closes the NSE by providing a relationship between the conserved variables and the pressure.
For a perfect gas, this can be written as
\begin{align}
  p &= \left(\gamma-1\right)\left(\rho e-\frac{\rho}{2}\:\ppvec{u}\cdot\ppvec{u}\right), \quad \text{or}\label{eq:eos_p}\\
  T &= \frac{p}{\rho R}.
  \label{eq:eos_T}
\end{align}
\Cref{eq:eos_p,eq:eos_T} thus allow the computation of the primitive, i.e.\ non-conserved, variables $\ppvec{U}^{prim}=(\rho,\ppvec{u},p,T)^T$ from the state $\ppvec{U}$.

\subsection{Discontinuous Galerkin Spectral Element Method}\label{sec:dgsem}

\begin{figure}[tb]
  \centering
  \includegraphics{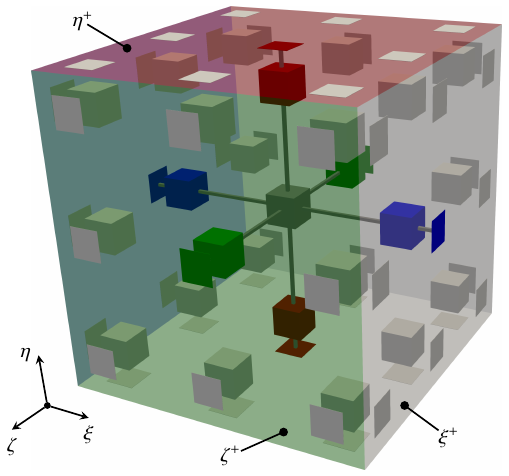}
   \caption{Perspective sketch of a single DG element in the reference space using Legendre-Gauss interpolation points with $N=2$. Gray cubes indicate the interpolation points within the element, while the gray squares indicate interpolation points on the six local faces called $\xi^{\pm},\eta^{\pm},\zeta^{\pm}$. The linewise operations of the tensor product ansatz are indicated for the center interpolation point, where the operations along the coordinates $\ppvec{\xi}=(\xi,\eta,\zeta)$ are highlighted in blue, red and green, respectively.}\label{fig:dg_cube}
\end{figure}

In the following section, the DGSEM will be derived for the compressible NSE, which can be written in flux formulation as
\begin{equation}
  \frac{\partial \ppvec{U}}{\partial t} + \nabla_{\!x}\cdot \ppvec{F}(\ppvec{U},\nabla_{\!x}\ppvec{U}) = \ppvec{0},
\end{equation}
where $\ppvec{F}(\ppvec{U},\nabla_{\!x}\ppvec{U})$ encapsulates both the convective and viscous fluxes.
Each of the main construction steps of the DGSEM will be discussed. However, a more in-depth derivation of the DGSEM and its implementation is provided by \citet{krais2021flexi}.

\subsubsection*{Mapping the Equations}
For the DGSEM, the domain $\Omega$ is subdivided into a set of non-overlapping, curvilinear, hexahedral elements.
Each physical element is then mapped from the physical space $\ppvec{x}=(x,y,z)^T$ to the reference element $E\in[-1,1]^3$ in computational space $\ppvec{\xi}=(\xi,\eta,\zeta)^T$ using a transfinite polynomial mapping $\ppvec{\xi}=\ppvec{\chi}(\ppvec{x})$.
The reference element is shown for $N=2$ in \cref{fig:dg_cube}.
The Jacobian~$\mathcal{J}$ of this mapping follows as the determinant of the Jacobian matrix $\nabla_{\!\xi}\,\ppvec{\chi}$, where $\nabla_{\!\xi}$ denotes the del operator in the computational coordinates.
The transformation of the governing equations into the computational space requires the contravariant basis vectors~$\mathcal{J}\!\ppvec{a}^i$, with $i=1,2,3$, which follow in the curl form as
\begin{equation}
  \mathcal{J}\!a^i_n = -\hat{x}_i\cdot \nabla_{\!\xi} \,\times\, (x_l \nabla_{\!\xi}x_m), \quad (n,m,l)\;\text{cyclic},
\end{equation}
where $\hat{x}_i$ is the unit vector in the $i$-th Cartesian direction.
Using the basis vectors and the Jacobian, the transformed equations in the reference element follow as
\begin{equation}
  \mathcal{J} \frac{\partial \ppvec{U}}{\partial t} + \nabla_{\!\xi}\cdot \ppvec{\mathcal{F}}^i = \ppvec{0},
  \label{eq:transformed_goveq}
\end{equation}
where $\ppvec{\mathcal{F}}^i$ denotes the contravariant fluxes given by
\begin{equation}
  \ppvec{\mathcal{F}}^i = \mathcal{J}\!\ppvec{a}^i\cdot \ppvec{F}.
\end{equation}
To construct the DGSEM, \cref{eq:transformed_goveq} is formulated in the weak form, which will be derived in the next paragraph.

\subsubsection*{Weak formulation}
To derive its weak form, \cref{eq:transformed_goveq} is projected onto a set of test functions $\psi(\ppvec{\xi})$, spanning a polynomial subspace, using the inner product, which yields
\begin{equation}
  \int\limits_E\! \mathcal{J} \frac{\partial \ppvec{U}}{\partial t} \psi \,\mathrm{d}\ppvec{\xi} + \underbrace{\oint\limits_{\partial E} \psi\,(\ppvec{\mathcal{F}}\cdot \ppvec{\mathcal{N}})^* \,\mathrm{d}S}_{\text{Surface Integral}} - \underbrace{\int\limits_E\! \ppvec{\mathcal{F}} \cdot  \nabla_{\!\xi}\psi \,\mathrm{d}\ppvec{\xi}}_{\text{Volume Integral}}= \ppvec{0}.
  \label{eq:weak_form}
\end{equation}
Here, the surface integral incorporates the contribution of the fluxes across the element faces while the volume integral considers only the degrees of freedom within the element.
Because adjacent elements share a common face and no continuity across elements has been imposed, the solution is generally discontinuous across the element faces.
Consequently, the solution and hence the fluxes on the element faces are non-unique.
Therefore, numerical flux functions are used to compute a unique numerical flux across element boundaries, which is denoted by the asterisk $(\cdot)^*$.
\subsubsection*{Solution Representation}
Within each element, the solution is represented by high-order Lagrange polynomials with any desired order resulting in arbitrary high-order accuracy of the DGSEM as demonstrated in \cref{sec:validation:convergence}.
The $j$-th one-dimensional Lagrange polynomial of degree $N$ is defined as
\begin{equation}
  \ell^N_i(x) = \underset{i\neq j}{\prod_{i=0}^N} \,\frac{x_i-x}{x_i-x_j},
\end{equation}
with respect to a set of interpolation points $\{x_j\}_{j=0}^N$.
In practice, either Legendre--Gauss~(GL) or Legendre--Gauss--Lobatto~(LGL) nodes are used as interpolation points.
The superscript is subsequently dropped to keep the notation concise.
Lagrange polynomials fulfill the Kronecker delta property given by
\begin{equation}
  \ell_i\left(x_j\right) = \begin{cases}
                  1, \quad\text{if}\;\; i=j,    \\
                  0, \quad\text{if}\;\; i\neq j.
                \end{cases}
\end{equation}
A tensor-product ansatz is used to construct a three-dimensional basis of the polynomial subspace $\mathbb{P}^N$ from the one-dimensional Lagrange polynomials.
This yields the approximation of the solution in the computational space as
\begin{equation}
  \ppvec{U}(\ppvec{\xi},t) \approx \sum_{i,j,k=0}^N \hat{\ppvec{U}}_{ijk}(t)\, \ell_i(\xi)\, \ell_j(\eta)\, \ell_k(\zeta).
  \label{eq:ansatz}
\end{equation}

\subsubsection*{Semi-discrete form}

Evaluating the integrals using the Gauss-type quadrature associated with the chosen set of interpolation points, i.e.\ collocation of interpolation and integration points,
yields the semi-discrete form of the DG operator that can be written for each point $i,j,k \in [0,N]$ as
\begin{alignat}{3}
  \frac{\partial \hat{\ppvec{U}}_{ijk}}{\partial t} = \overbrace{-\frac{1}{\mathcal{J}_{ijk}}}^{\text{\code{ApplyJac}}}\Bigg[&
       \sum_{\alpha=0}^N \ppvec{\mathcal{F}}^1_{\!\!\! \alpha jk} \hat{D}_{i \alpha} &&+ \Big( \overbrace{\left[\ppvec{f}^*\hat{s}\right]_{jk}^{ \xi^+}}^{\text{\code{FillFlux}}} \hat{\ell}^+_i + \overbrace{\left[\ppvec{f}^*\hat{s}\right]_{jk}^{ \xi^{-}}}^{\text{\code{FillFlux}}} \hat{\ell}^-_i \Big)\nonumber \\
    +& \sum_{\beta =0}^N \ppvec{\mathcal{F}}^2_{\!\!\! i \beta k} \hat{D}_{j \beta } &&+ \Big( \left[\ppvec{f}^*\hat{s}\right]_{ik}^{\eta^+} \hat{\ell}^+_j + \left[\ppvec{f}^*\hat{s}\right]_{ik}^{\eta^{-}} \hat{\ell}^-_j \Big)\nonumber \\
    +& \underbrace{\sum_{\gamma=0}^N \ppvec{\mathcal{F}}^3_{\!\!\! i j \gamma} \hat{D}_{k \gamma}}_{\text{\code{VolInt}}}  &&+ \underbrace{\Big(     \left[ \ppvec{f}^*  \hat{s}
    \right]_{ij}^{\zeta^+} \hat{\ell}^+_k + \left[ \ppvec{f}^*  \hat{s} \right]_{ij}^{\zeta^{-}} \hat{\ell}^-_k   \Big)\vphantom{\sum_{\gamma=0}^N}}_{\text{\code{SurfInt}}}
    \Bigg].
  \label{eq:dgsem}
\end{alignat}
This notation follows \citet{kopriva2009implementing}, where
\begin{equation}
\hat{\ell}_i^{\pm} = \frac{\ell_i(\pm 1)}{\omega_i} \qquad\text{and}\qquad \hat{D}_{ij} = -\frac{\omega_i}{\omega_j}\left.\frac{\mathrm{d}\ell_i(\xi)}{\mathrm{d}\xi}\right|_{\xi=\xi_j} \end{equation}
are one-dimensional building blocks that entail the numerical quadrature weights $\omega_i$ and are precomputed during initialization to improve the overall performance of the implementation.
Moreover, $\ppvec{f}^*=\ppvec{f}^*(\tilde{\ppvec{U}}^L,\tilde{\ppvec{U}}^R)$ denotes the unique flux at the faces based on the solution on the surface of the left and right element, respectively, and $\hat{s}$ denotes the surface element, which is the norm of the non-normalized physical unit vector as discussed in more detail by \citet{krais2021flexi}. The monospaced namings in \cref{eq:dgsem} refer to the routines in the numerical implementation which are summarized in \cref{tab:dg_routines} and are also detailed in \ref{app:dg_routines}.

At this point, we would like to briefly discuss the influence of the unstructured mesh topology. First of all, the unstructured neighbor relations only influence the surface-related operations and only direct neighbors are considered. Here, the relative orientation between the adjacent elements and their faces must be taken into account. This is taken into account by corresponding mappings in the \code{ProlongToFace} and \code{SurfInt} routines.

\begin{table*}
  \centering
  \begin{tabularx}{\textwidth}{lccccX}
    \toprule
    Routine              & Vol/Surf            & DOF-local & \code{Lift\_*} & Operations             & Explanation\\
    \midrule
    \code{ConsToPrim   } & Surf, Vol           &       YES & NO             & $\mathcal{O}(N^{2,3})$ & Computes primitive variables $\ppvec{U}^{prim}$ from state $\ppvec{U}$.\\
    \code{VolInt       } & Vol                 &       NO  & YES            & $\mathcal{O}(N^4)$     & Evaluates volume fluxes $\ppvec{\mathcal{F}}$ and multiplies with $\hat{\ppvec{D}}$.\\
    \code{ProlongToFace} & Vol$\rightarrow$Surf&       NO  & YES            & $\mathcal{O}(N^2)$     & Evaluates solution at element faces $\ppvec{U}^{L/R}$ to compute $\ppvec{f}^*$.\\
    \code{FillFlux     } & Surf                &       YES & YES            & $\mathcal{O}(N^2)$     & Computes common flux $\ppvec{f}^*$ on faces with Riemann solver.\\
    \code{SurfInt      } & Vol$\leftarrow$Surf &       NO  & YES            & $\mathcal{O}(N^2)$     & Computes surface integral with $\ppvec{f}^*$ and $\hat{\ppvec{\ell}}^{\pm}$.\\
    \code{ApplyJac     } & Vol                 &       YES & YES            & $\mathcal{O}(N^3)$     & Applies Jacobian $\mathcal{J}$ to $\hat{\ppvec{U}}_t$.\\
    \bottomrule
  \end{tabularx}
  \caption{Individual operations required to evaluate the three-dimensional DG operator with details on whether the routine acts on volume data or surface data. Moreover, it is indicated whether performed operations are DOF-local, i.e.\ are performed independently for each specific DOF, if they have to be re-applied during the computation of the gradients, which is indicated by the prefix $\code{Lift\_*}$, and their computational complexity in terms of $N$. A more detailed discussion of these routines can be found in \ref{app:dg_routines}.}\label{tab:dg_routines}
\end{table*}

\subsubsection*{Time integration}
The semi-discrete form \cref{eq:dgsem} is integrated in time using an appropriate integration scheme.
\galexi offers a variety of different explicit high-order Runge--Kutta-type schemes in a low-storage formulation to reduce the memory consumption.
In the following, a fourth-order Runge--Kutta scheme with 5 stages~\cite{carpenter1994rk} is used for the validation and verification results in \cref{sec:validation}, while a scheme with 14 stages~\cite{niegemann2012efficient} is used for the large-scale application in \cref{sec:application}. The latter scheme is chosen since it exhibits an optimized stability region for convection-dominated problems and allows for larger time steps.

\subsubsection*{Nonlinear Stability}
The semi-discrete form of the DG discretization in \cref{eq:dgsem} is derived by means of numerical quadrature rules.
However, since the integrands, i.e.\ the fluxes of the compressible NSE, are non-polynomial, they cannot be integrated exactly by the applied quadrature rules.
The resulting integration errors manifest as aliasing that can cause simulations to crash, especially in the underresolved regime.
For DG, multiple mitigation strategies have been devised ranging from overintegration~\cite{kirby2003aliasing,beck2016influence}, also referred to as polynomial dealiasing, to filtering procedures that strive to counteract the accumulation of energy in the highest solution modes~\cite{hesthaven2007nodal,flad2016simulation}.
In this work, we rely on the split-flux formulation introduced by \citet{gassner2016split} to construct a non-linear stable DG scheme.
This approach is based on the strong formulation of the governing equations, which can be obtained through a second integration-by-parts of \cref{eq:weak_form}.
The discretized form can be cast into the same algorithmic form as \cref{eq:dgsem} with only minor modifications in the formulation of the fluxes~\cite{krais2021flexi}.
Here, the fluxes of the NSE are replaced by split-form two-point fluxes that are equivalent on an analytical level but can be used to enforce additional constraints such as entropy consistency in the discretized formulation.
In this work, a kinetic-energy-preserving split-flux formulation proposed by \citet{pirozzoli2011numerical} is applied.
It is important to stress that the evaluation of two-point fluxes increases the computational cost considerably.

\subsubsection*{Second-Order Equations}
For the NSE, the gradients of the primitive variables $\nabla_{\!x} \ppvec{U}^{prim}$ are required to evaluate the viscous fluxes.
While several approaches exist in the literature, \galexi follows the \textit{BR1} method by \citet{bassi1997high}.
Here, so-called \emph{lifted} gradients $\ppvec{g}$ are introduced that should fulfill
\begin{equation}
  \ppvec{g}-\frac{1}{\mathcal{J}}\,\nabla_{\!x} \ppvec{U}^{prim}=\ppvec{0}.
  \label{eq:lifting}
\end{equation}
This equation is then solved for $\ppvec{g}$ by deriving the weak form of \cref{eq:lifting} and applying the DGSEM as is done for the NSE themselves.
This yields an additional set of equations that is structurally similar to \cref{eq:dgsem} but using the lifting fluxes instead of the fluxes of the NSE as detailed in \citet{krais2021flexi}.
Hence, the computation of the gradients corresponds to increasing the set of unknowns by an additional $(n_{\text{dim}}\times n_{\text{lift}})$ variables, where $n_{\text{dim}}$ corresponds to the number of spatial dimensions and $n_{\text{lift}}$ to the number of primitive variables for which the gradients should be computed.
This also means that each of the operations indicated in \cref{eq:dgsem}, i.e.\ \code{ApplyJac}, \code{SurfInt}, \code{VolInt}, \code{FillFlux}, has to be executed again for the lifting procedure in each spatial direction.

\subsubsection*{Computational Complexity}
\Cref{tab:dg_routines} also provides the estimated number of operations, i.e.\ the computational complexity, for the different steps of the three-dimensional DG discretization.
The given numbers describe the asymptotic behavior of the individual operations without considering details such as the computational complexity of the flux computation and compiler optimizations.
Most importantly, the computational effort to compute the volume integral scales one order higher in terms of $N$ than all other operations, i.e.\ it scales with $\mathcal{O}(N^4)$ instead of $\mathcal{O}(N^3)$ or even $\mathcal{O}(N^2)$.
Hence, the volume integral becomes the dominant operation for increasing $N$.
However, the computations carried out for the volume integral are purely element local, highly dense and can be computed very efficiently on various types of hardware.
In practice, this increase in efficiency was observed to partly compensate for the additional operations required with increasing $N$~\cite{beck2014high}.
Moreover, the communication stencil between elements is small, since only surface fluxes with direct neighbor elements have to be exchanged.
Consequently, the high cost of the volume integral and the small communication stencil allow for hiding the communication latencies very efficiently in parallel computations.

\subsection{Shock Capturing}\label{sec:shock_capturing}

\begin{figure}[t]
  \includegraphics{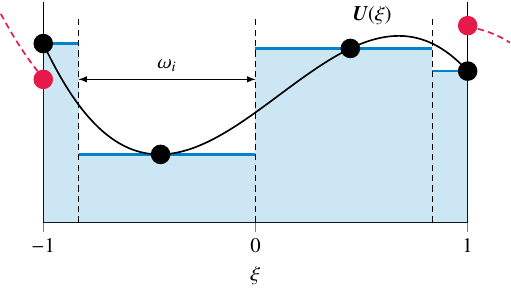}
   \caption{Sketch of the sub-cell shock capturing scheme. The DG polynomial using LGL points and a polynomial degree of $N=3$ is shown in black with the interpolation points indicated as dots and the integral mean solution within the subcells is shown in blue. The solution in the neighboring DG elements is indicated in red.}
  \label{fig:dgfv}
\end{figure}

\galexi is designed for the simulation of compressible flows which can entail discontinuities in the form of shocks.
However, the application of high-order discretizations near discontinuities or strong gradients in the solution produces spurious oscillations and can cause the numerical scheme to become unstable.
As a consequence, a wide variety of different shock identification and capturing methods are proposed in the literature that all strive to stabilize high-order discretizations near shocks and provide stable and accurate simulations of compressible flows.
The common objective of those methods is to retain the high-order accuracy of the baseline scheme in smooth regions while identifying and handling so-called troubled cells within the domain during the simulation.
Already \citet{cockburn1989tvb} used limiters to stabilize DG methods near shocks, which was in parts inspired by the total variation diminishing/bounded properties of FV schemes.
Based on this, a myriad of different limiting techniques have been proposed in the literature~\cite{krivodonova2007limiters,zhang2010positivity,burbeau2001problem}.
Other common approaches introduce some form of artificial viscosity near the shock region~\cite{zeifang2021data,persson2006sub,klockner2011viscous} or use high-order filtering techniques~\cite{bohm2019multi}.
Another approach is to employ a hybrid discretization, where the high-order DG scheme is stabilized in the vicinity of the troubled region with a low-order FV scheme.
For this, the DG element is subdivided into multiple FV subcells as indicated in \cref{fig:dgfv}.
This low-order scheme can then either be solved directly within the troubled elements and coupled to the surrounding DG elements using the common Riemann fluxes~\cite{sonntag2014shock} or can also be used as a regularizing limiter~\cite{fambri2017space}.
In the following, we employ the blending approach by \citet{hennemann2021provably}, who proposed to compute a convex blending of both discretization operators. This approach has also been demonstrated to yield a sensible turbulence model if tuned correctly~\cite{beck2023toward}.
Within each element both the high-order DG operator $\ppvec{\mathcal{R}}^{\text{DG}}(\hat{\ppvec{U}})$ and the compatible low-order FV scheme $\ppvec{\mathcal{R}}^{\text{FV}}(\hat{\ppvec{U}})$ are evaluated.
The convex blending of both schemes then yields
\begin{equation}
  \frac{\partial \hat{\ppvec{U}}}{\partial t} = (1-\alpha)\,\ppvec{\mathcal{R}}^{\text{DG}}(\hat{\ppvec{U}}) + \alpha\,\ppvec{\mathcal{R}}^{\text{FV}}(\hat{\ppvec{U}}),
\end{equation}
where the blending factor $\alpha\in[0,1]$ can be computed either via an \emph{a priori} or \emph{a posteriori} strategy for each individual DG element~\cite{rueda2022subcell}.
In this work, the \emph{a priori} indicator by \citet{hennemann2021provably} is used, for which the blending approach becomes an operator local to each individual element.
Clearly, the standard DG scheme can be recovered for $\alpha=0$, while $\alpha=1$ yields a pure FV discretization.
It is important to stress that only the contributions of the operators within the element have to be blended since the outer surface fluxes are identical for the DG and FV formulation.

\section{Parallelization Strategy on Accelerators}\label{sec:implementation}

\galexi is the endeavor to extend the flow solver FLEXI~\cite{krais2021flexi} towards accelerator-based HPC systems.
Here, \galexi follows three distinct design principles.
First, the general data structure and parallelization strategy of FLEXI for unstructured geometries should be retained.
Second, we strive to retain the majority of the codebase and the associated features of the original implementation.
Lastly, \galexi is designed such that all routines called during the time-stepping are executed on the accelerator without the need to transfer data to and from the CPU.
Device code and compute kernels are only required for routines that are called during time-stepping and thus have to be computed on the accelerator.
In contrast, initialization and non-frequently performed analyzing routines are still computed on the CPU, since they are less time-critical and CPUs are better suited towards unstructured workloads.
Both \galexi and FLEXI are implemented in modern Fortran 2008.
The device code for the accelerators in \galexi is currently implemented using CUDA Fortran, but the integration of other compute backends is under development.

The design and implementation of \galexi is detailed in the following sections using a hierarchical top-down approach.
First, the high-level distribution of work across different compute devices and the employed communication scheme between them is detailed in \cref{sec:implementation_multigpu}.
Based on this, \cref{sec:implementation_gpu} provides details on how communication and compute kernels are arranged and scheduled on a single GPU.
Lastly, the general implementation par\-a\-digms for the individual compute kernels are detailed in \cref{sec:implementation_kernel}.
Obviously, performance optimizations have to be performed across all of these three levels and changes on one level affect the suitability and performance of the others.
While these individual levels are inherently interlinked, we chose this partitioning in the following to provide a more structured over\-view of the design principles and methods applied in \galexi.

\subsection{Inter-GPU Parallelization}\label{sec:implementation_multigpu}

\begin{figure*}
  \centering
  \includegraphics[width=0.9\textwidth]{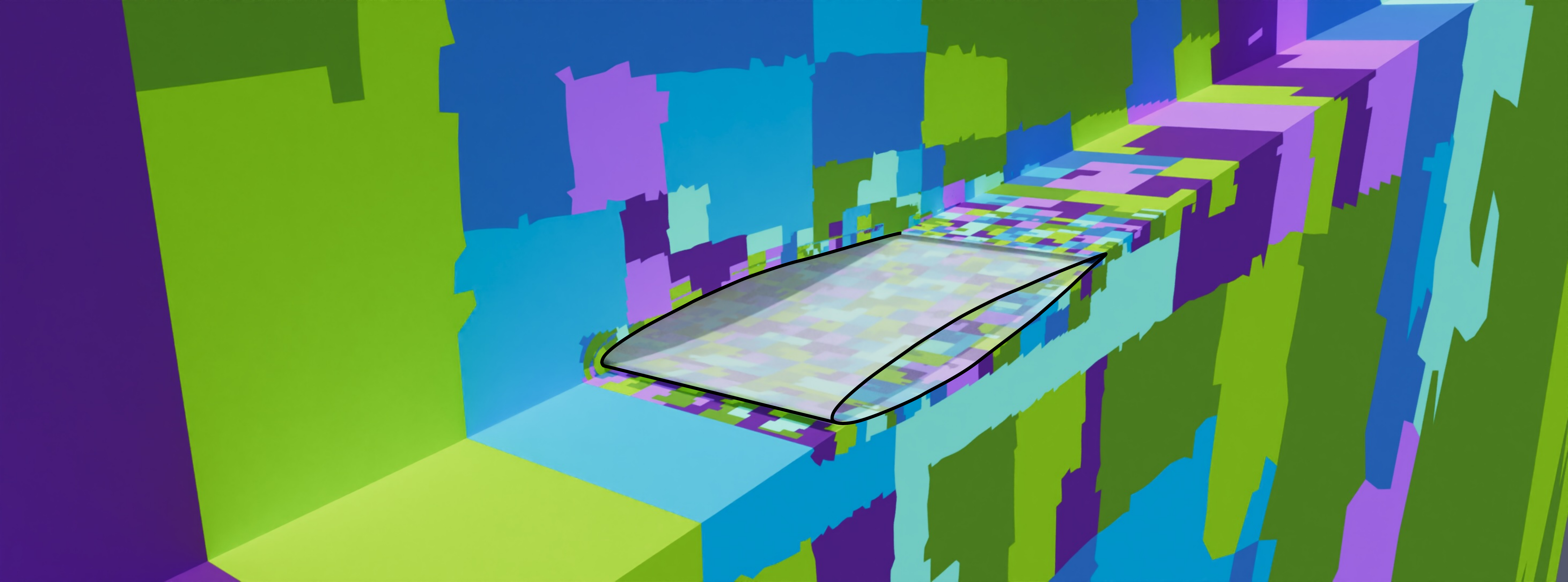}
  \caption{Domain decomposition for a generic airfoil simulation with large spanwise extension. The domain is cut such that the airfoil (transparent surface) including the boundary layer part is visible. Patches of different colors represent individual MPI domains that are processed by different ranks. This figure is an example of a fine granularity, e.g.\ in the CPU case. In the GPU case, larger MPI domains occur.}\label{fig:domain_decomposition}
\end{figure*}

The parallelization strategy between GPUs in \galexi is large\-ly inherited from FLEXI, which employs a pure distributed memory approach using MPI.
Before going into the specific implementation details of \galexi, the original MPI parallelization strategy of FLEXI is briefly presented.
Here, each computational rank is assigned a subdomain of roughly the same amount of elements as shown in \cref{fig:domain_decomposition}.
In the DG context, elements are only coupled via their surface fluxes.
Thus, only the surface information across the MPI borders has to be exchanged between individual MPI ranks during the computation.
Moreover, FLEXI sorts this face information such that the data exchanged between two MPI partners is contiguous in memory and that the sorting is known a priori on both faces.
This allows to exchange solely the data itself without any additional sorting information.
The overall communication effort is thus proportional to the number of faces at the MPI boundaries, which are referred to in the following simply as \textit{MPI faces}.
To minimize the amount of communication, i.e.\ the number of MPI faces in the domain, FLEXI distributes the domain along a pre-computed space-filling curve.
This ensures that the resulting subdomains remain reasonably compact for any amount of subdomains while minimizing partitioning effort. During the simulation, communication is generally asynchronous and non-blocking, which means that communicated data is computed and sent at the earliest possible opportunity.
The communication barrier that checks whether the data has been received is positioned at the latest possible instant before the data is actually required for further computations.
This allows to effectively hide the communication latency by performing local operations during the data exchange.
For this, operations on MPI faces are prioritized over inner faces to use all operations performed on inner faces for latency hiding.

\galexi follows the same general approach for parallelizing across multiple GPUs.
Here, each GPU on a node is associated with a distinct CPU core while respecting the memory topology to optimize performance.\footnote{This means, for instance, to associate the CPU core and the GPU such that both reside within the same non-uniform memory access (NUMA) domain.}
Moreover, CUDA-aware\footnote{For further information about CUDA-aware MPI see: \url{https://docs.open-mpi.org/en/v5.0.x/tuning-apps/networking/cuda.html}} implementations of MPI are used, which improve the performance of MPI communication.
By providing Remote Direct Memory Access (RDMA) and host offloading, these allow direct access to the local memory of different GPUs on the same node and transmission of MPI messages directly from the GPU to the network adapter without the assistance of the CPU or the main memory.
The following key differences emerge between the CPU and GPU implementations.
First, the domain size on a single GPU is larger than for the CPU case.
This is a result of the much higher computational power a GPU provides compared to a single CPU core.
A GPU requires significantly more workload to run at capacity and to exploit the full degree of its parallelism.
In practice, this means that the computational domain per rank increases if GPUs are used.
Since the subdomains are compact, an increase in size means that the volume increases much faster than the MPI surface, i.e.\ that inner work becomes more dominant in comparison to the required communication and at the same time the amount of data to be communicated decreases. In consequence, the performance of the interconnect becomes less dominant than in the CPU case.
Second, the GPU implementation has to consider the asynchronicity between the GPU device and the host.
While an operation is launched by the host at a specific position in the code sequence, the GPU schedules and executes the operation independently of the work performed by the host in the meantime.
Hence, additional synchronization between the host and the device is necessary to ensure data consistency.
This entails for instance ensuring that a buffer that is about to be sent via MPI has already been filled with the required information by the GPU.
This introduces additional overhead. However, due to the asynchronous operation, CPU and device operations can again be overlapped, which results in an additional level of parallelism on an intra-GPU level and is addressed in \cref{sec:implementation_gpu}.

\subsection{Intra-GPU Parallelization}\label{sec:implementation_gpu}

\begin{figure}[tb]
  \centering
  \includegraphics{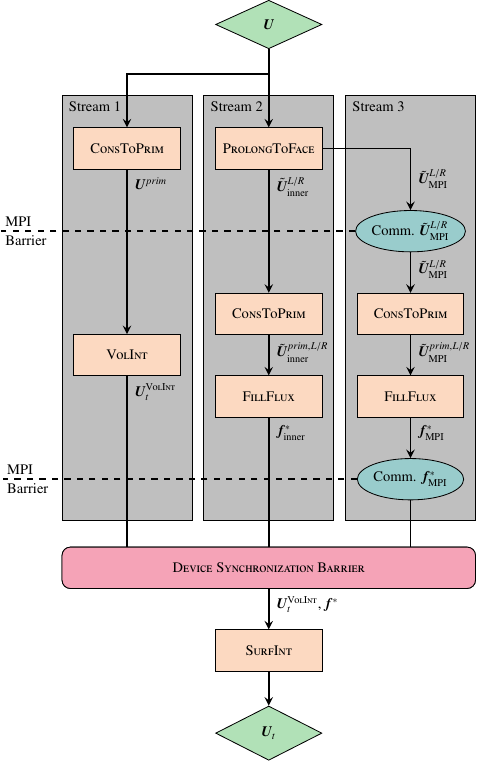}
   \caption{Flowchart of \galexi for a single evaluation of the convective DG operator using streams. Some routines comprise several individual compute kernels instead of single, monolithic device kernels. These are summarized here to keep the flowchart concise. Moreover, the lifting procedure to compute the gradients is omitted here for readability.}\label{fig:flowchart}
\end{figure}

As already discussed in the previous paragraphs, device kernels are launched within host code.
However, the GPU schedules and executes the launched kernels asynchronously and can also execute multiple kernels concurrently to maximize its utilization.
It is important to consider these properties to maximize the achieved performance on the device.
\galexi relies on so-called \emph{streams} to manage the concurrency and scheduling of operations on the GPU.
Within the GPU context, streams are similar to execution pipelines.
Kernels within each stream are executed serially, i.e.\ the next kernel within a stream pipeline is only executed once all preceding kernels within this distinct pipeline have finished execution.
However, kernels from different streams can run concurrently on the GPU to maximize utilization.
This can improve the overall performance either when running small kernels that cannot fully utilize the GPU or by hiding the overhead associated with starting a kernel on the device.
Another benefit is that streams allow the mitigation of the \emph{tail effect}, which describes the negative performance impact of the last partial wave of computations in a kernel.
This effect stems from the last thread blocks of a kernel call which will generally not fill the whole GPU, leading to a significant portion of the GPU idling when the last wave of computations are performed.
By using streams, the idling resources can execute kernels from different streams that are known to be independent of the current computation, which improves GPU utilization and thus the overall performance.

A key factor when using streams is to ensure correct results independent of how the individual kernels are scheduled.
Therefore, another level of synchronization between the streams is required to mitigate race conditions.
In \galexi, the different operations of the convective DG operator, summarized in \cref{tab:dg_routines}, are assigned to individual streams depending on their interdependence.
This means that if one kernel requires a previous kernel to be completed, both are assigned to the same stream to be executed sequentially.
In contrast, operations that are independent of each other get assigned to different streams.
In \galexi, three streams are employed to account for the available concurrency:
\begin{itemize}
  \item \textbf{Stream \makebox[\widthof{3}][r]{1}} (priority \makebox[\widthof{mid}][r]{low}): \makebox[\widthof{Operations within DG elements.}][l]{Operations within DG elements.}
  \item \textbf{Stream \makebox[\widthof{3}][r]{2}} (priority \makebox[\widthof{mid}][r]{mid}): \makebox[\widthof{Operations within DG elements.}][l]{Operations on inner faces.}
  \item \textbf{Stream \makebox[\widthof{3}][r]{3}} (priority \makebox[\widthof{mid}][r]{top}): \makebox[\widthof{Operations within DG elements.}][l]{Operations on MPI faces.}
\end{itemize}
Here, each stream is assigned a priority which incentivizes the GPU to preempt and postpone the execution of low-priority kernels in favor of high-priority ones.
In \galexi, Stream 3 containing the MPI faces is assigned the highest priority to ensure that data that has to be communicated is always computed at the earliest possible instant to ensure optimal latency hiding.
The flowchart for the evaluation of the DG operator using these streams is shown in \cref{fig:flowchart}.
The local operations within the DG element and the operations at the element faces can be performed independently in their streams until the computation of \code{SurfInt}, where the surface fluxes $\ppvec{f}^*$ as well as the contributions of the \code{VolInt} denoted $\ppvec{U}_t^{\text{\code{VolInt}}}$ are required.
Hence, an explicit synchronization barrier is employed to wait until all previous operations in all streams have completed.
Then, the surface contributions can be added to the volume integral to yield the final $\ppvec{U}_t$.

Special care must be taken when extending the operator towards multiple GPUs, which requires MPI communication.
Here, it has to be ensured that all kernels \textit{within} the GPU are assigned correctly to individual streams and that the MPI communication \textit{between} GPUs is effectively hidden by the local work. For this, the work associated with MPI faces, i.e.\ Stream 3, is ensured to be computed with the highest priority, such that the communication can be initialized as soon as possible.
The work queued in the other streams is then used to hide both the local overhead of tail effects on the GPU and the latency of the MPI communication.
In practical application, the host idles at the MPI barrier until the communication is finished, but the GPU is kept busy with the work from Stream 1 and Stream 2 to retain the overall efficiency.
Effectively, the communication of the solution at the MPI faces $\tilde{\ppvec{U}}^{L/R}_{\text{MPI}}$ is hidden by \code{ConsToPrim} on Stream 1.
The communication of the resulting fluxes across the MPI faces $\ppvec{f}^*_{\text{MPI}}$ is hidden by \code{VolInt} in Stream 1 and \code{ConsToPrim} and \code{FillFlux} in Stream 2.

\subsection{Kernel Implementation}\label{sec:implementation_kernel}

The goal for the implementation of the compute kernels is to maximize the utilization of the available parallel resources provided by the device which roughly translates to keeping as many threads as possible busy.
However, oftentimes the number of concurrent threads is limited by the number of registers required by each thread and the amount of shared memory.
Futhermore, the effective performance is limited by the available memory bandwidth, which might not be sufficient to keep all threads busy.

Since the specifics of these limitations and their importance depend heavily on the specific hardware, \galexi approaches the problem from a more general perspective.
Here, it is assumed that addressing these GPU-specific limitations in general and improving the overall performance for some generic GPU architecture also yields sensible improvements across all specific GPU models.
While this approach may not achieve the optimum performance for each specific type of hardware, it has demonstrated to yield a suitable baseline for more in-depth and hardware-specific optimization.

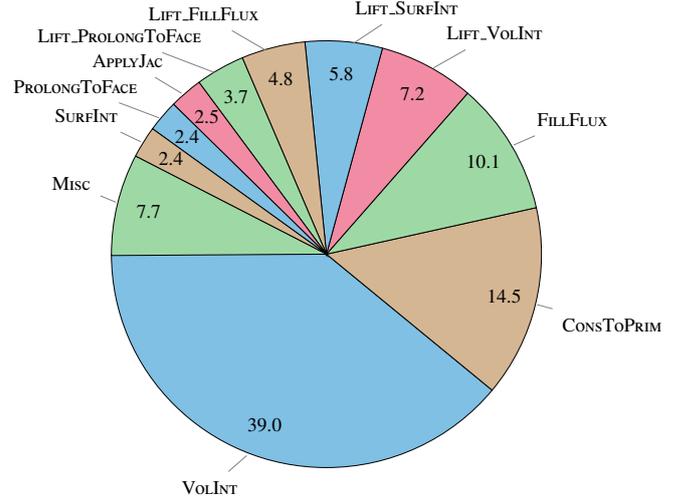
\begin{figure}[tb]
  \centering
  \begin{tikzpicture}[nodes = {font=\scriptsize}]

\def\angle{180}
  \def\pindistance{0.2cm}
  \def\radius{2.7}
  \def\cyclelist{{"colora","colorb","colorc","colord"}}
  \newcount\cyclecount \cyclecount=-1
  \newcount\ind \ind=-1

  \foreach \percent/\name in {
      39.0/\code{VolInt},   14.5/\code{ConsToPrim}, 10.1/\code{FillFlux}, 7.2/\code{Lift\_VolInt},
       5.8/\code{Lift\_SurfInt},
       4.8/\code{Lift\_FillFlux},
       3.7/\code{Lift\_ProlongToFace},
       2.5/\code{ApplyJac},
       2.4/\code{ProlongToFace},
       2.4/\code{SurfInt},
7.7/\code{Misc} } {
      \ifx\percent\empty\else               \global\advance\cyclecount by 1     \global\advance\ind by 1            \ifnum3<\cyclecount                 \global\cyclecount=0              \global\ind=0                     \fi
        \pgfmathparse{\cyclelist[\the\ind]} \edef\color{\pgfmathresult}         \draw[fill={\color!50},draw={black},ultra thin] (0,0) -- (\angle:\radius)
          arc (\angle:\angle+\percent*3.6:\radius) -- cycle;
        \node at (\angle+0.5*\percent*3.6:0.85*\radius) {\percent};
        \node[pin={[pin distance=\pindistance]\angle+0.5*\percent*3.6:\scriptsize\name}]
          at (\angle+0.5*\percent*3.6:0.96*\radius) {};
        \pgfmathparse{\angle+\percent*3.6}  \xdef\angle{\pgfmathresult}         \fi
    };
\end{tikzpicture}
   \caption{Portion of compute time in percent for individual routines on HAWK-AI with $N=7$, split-form DG and \SI{8.9e5}{DOF} on a single GPU. Routines associated with the computation of the gradients via the lifting method are prefixed with ``\code{Lift\_}''. Various small routines associated with performing the actual time integration, i.e.\ updating $\ppvec{U}$, are summarized under \code{Misc}.}\label{fig:galexi_piechart}
\end{figure}

Device code is typically based on a \emph{kernel}, which is the code each individual thread executes.
The overall number of threads and their grouping are specified in the \emph{launch configuration}.
In some sense, the launch configuration entails an implicit tightly nested loop, while the loop body, i.e.\ the actual computation, is implemented in the kernel.
The optimal launch configuration is oftentimes highly hardware-specific and can improve (or impair) the overall performance significantly.
Along the same lines as discussed above, our code relies on sensible initial guesses for all of these kernels, which gave reasonable results.
Further improvements are planned through the application of more sophisticated tuning approaches, for instance the \emph{kernel tuner} toolkit~\cite{kerneltuner}, which allows automatized optimization of the launch configuration for specific hardware.

The complete list of operations of the DG operator is detailed in \cref{tab:dg_routines} and the computing time of the kernels associated with these operations is summarized in \cref{fig:galexi_piechart}.
Naturally, operations that are DOF-local are the easiest to implement for different hardware.
Hence, the following paragraph first introduces how kernels are designed for DOF-wise operations in \galexi before moving to the much more intricate task of kernels that map data between the volume and faces of DG elements.

\subsubsection*{Pointwise operations}

\begin{algorithm}[tb]
  \caption{Wrapper for \code{ConsToPrim\_Point} on CPU}\label{alg:constoprim_cpu}
  \begin{algorithmic}[1]
    \Function{ConsToPrim\_CPU}{$N,n_{Elems},\ppvec{U}$}
      \For{$n \gets 1\text{ to }n_{Elems}$} \Comment{loop \makebox[\widthof{within element}][r]{over elements}}
        \For{$i,j,k \gets 0\text{ to }N$}   \Comment{loop within element}
          \State $\ppvec{U}^{prim}_{ijk,n} \gets \Call {ConsToPrim\_Point}{\ppvec{U}_{ijk,n}}$
        \EndFor
      \EndFor
      \State \Return {$\ppvec{U}^{prim}$}
    \EndFunction
  \end{algorithmic}
\end{algorithm}

\begin{algorithm}[tb]
  \caption{Wrapper for \code{ConsToPrim\_Point} on GPU}\label{alg:constoprim_gpu}
  \begin{algorithmic}[1]
    \Function{ConsToPrim\_GPU}{$N,n_{Elems},\ppvec{U}$}
        \State{$n_{DOF} \gets (N+1)^3 n_{Elems}$ }    \Comment{number of DOF in array}
        \State $\ppvec{U}^{prim} \gets \Call {ConsToPrim\_Kernel<<config>>}{n_{DOF},\ppvec{U}}$
        \State \Return {$\ppvec{U}^{prim}$}
    \EndFunction
    \State \Kernel{ConsToPrim\_Kernel}{$n_{DOF},\ppvec{U}$}
      \State{$i \gets $\code{(blockID-1)*blockDim+threadID}} \Comment{own index}\label{code:config}
      \If{$i \leq n_{DOF}$}
        \State $\ppvec{U}^{Prim}_{i} \gets \Call {ConsToPrim\_Point}{\ppvec{U}_{i}}$
        \State \Return {$\ppvec{U}^{prim}_i$}
      \EndIf
    \EndKernel
  \end{algorithmic}
\end{algorithm}

For pointwise operations, a large number of identical computations have to be performed with no interdependence between individual DOFs.
Such a computation becomes embarrassingly parallel and straight\-for\-ward to distribute.
The following paragraph details how such computations are implemented in \galexi using the \code{ConsToPrim} operation as an example.
This operation computes the primitive variables $\ppvec{U}^{prim}=(\rho,\ppvec{u},p,T)^T$ based on the vector of conservative variables $\ppvec{U}$ using the EOS defined in \cref{eq:eos_p,eq:eos_T}.
For this, an elemental \code{ConsToPrim\_Point} routine is implemented that performs the computation for a single DOF.
This elemental routine is the building block of the main computation and is agnostic to the underlying hardware.
\galexi then uses different wrappers for this elemental function.
These wrappers distribute the overall work depending on the specific type of computational hardware used.
If CPUs are used, the design of the wrapper becomes straightforward as shown in \cref{alg:constoprim_cpu}.
A single CPU core just calls the \code{ConsToPrim\_Point} routine for each DOF within each element of its domain using a tightly nested loop.
The GPU wrapper shown in \cref{alg:constoprim_gpu} is based on the CUDA programming model and consists of two individual components.
First, the \textit{kernel} that implements the actual compute operation of an individual GPU thread.
The second component is a function that calls the kernel and provides the \textit{launch configuration} \code{config}.
The launch configuration determines how many threads will be started to execute the kernel and how the individual threads are grouped into thread blocks.
In this specific case, each thread of the GPU performs the computation for a single DOF in the domain.
For this, each thread determines in line~\ref{code:config} of \cref{alg:constoprim_gpu} its own globally unique thread ID $i$.
This thread ID incorporates the ID of the current block (\code{blockID}), the size of each block (\code{blockDim}), and its thread number within the block (\code{threadID}), which are all available for each thread during runtime.
The thread then performs the computation for this $i$-th DOF.
Note that the high-dimensional structure of the array becomes irrelevant in this case and can be ``flattened'' to a one-dimensional array containing $n_{DOF}$ entries.

More advanced techniques can be used to optimize those wrappers for different hardware.
This includes for instance vectorization, such that either the vector units of a CPU or real vector accelerators can perform the operations performed in \code{ConsToPrim} on several entries of $\ppvec{U}$ simultaneously.
Similarly, optimization such as loop unrolling or shared memory parallelization are straightforward to implement.
For GPU usage, the wrapper can be adapted to distribute multiple DOFs to each thread and optimize the launch configuration, depending on the hardware specifics.
The same building block approach can also be applied to support other backends such as HIP, ROCm, OpenMP or OpenACC while only maintaining a single version of the equation-specific code.

\subsubsection*{Volume$\leftrightarrow$ Surface Operations}
The optimization potential of the pointwise operations discussed above is mostly independent of the core algorithms themselves.
In contrast, the most challenging routines for GPU porting and parallelization in the DG context are routines that map data between the faces and the volume.
Due to the highly local nature of the DG method, the transfer of data from within the element to its faces and vice versa is required in only a few operations.
Thus, the original FLEXI code opted to store the data on the element faces and within the elements in different arrays from which it is retrieved based on precomputed mappings.
However, revisiting \cref{tab:dg_routines} reveals that two specific operations in the DG operator access both volume and surface data: \code{ProlongToFace} and \code{SurfInt}.\footnote{As shown in \cref{tab:dg_routines}, the volume integral is also not a point-local operation due to the application of the differentiation matrix along the lines indicated in \cref{fig:dg_cube}.
However, the operations are retained to the interpolation points within the volume of the DG element, i.e.\ no exchange of information between the volume and the faces is required.}
The former evaluates the polynomial solution from the interior points at the element faces and stores it in a face-based array ($\ppvec{U}\rightarrow\tilde{\ppvec{U}}^{L/R}$), while the latter computes the integral of the fluxes on the element faces and adds their contribution to the volume ($\ppvec{f}^*\rightarrow\ppvec{U}_t$).
In both cases, an interpolation point in the volume is linked to several points on the surface and vice versa, as shown in \cref{fig:dg_cube}.
Special care must be taken to exploit the full potential for parallelization of the task on a GPU while avoiding race conditions and costly synchronizations among individual threads. In the following, this is illustrated for the \code{SurfInt} routine.

\begin{algorithm}[tbh!]
  \caption{CPU implementation of the \code{SurfInt} operation}\label{alg:surfint_cpu}
    \begin{algorithmic}[1]
    \Function{SurfInt}{$\ppvec{f}^{*},\ppvec{U}_t,\hat{\ppvec{\ell}}^+,\hat{\ppvec{\ell}}^-$}
    \For{$s \gets 1\text{ to }n_{Faces}$}
\If{isPrimary}
\State $\ppvec{f}^{*,\text{tmp}},\code{locFace} \gets \Call{FaceMapping}{s,\code{isPrimary},\ppvec{f}^{*}_{s}}$
        \State $\ppvec{U}_{t} \gets \Call{DoSurfInt}{\code{locFace},\ppvec{U}_t,\ppvec{f}^{*,\text{tmp}},\hat{\ppvec{\ell}}^+,\hat{\ppvec{\ell}}^-} $
      \EndIf
      \If{\code{isReplica}}
\State $\ppvec{f}^{*,\text{tmp}},\code{locFace} \gets \Call{FaceMapping}{s,\code{isPrimary},-\ppvec{f}^{*}_{s}}$
        \State $\ppvec{U}_{t} \gets \Call{DoSurfInt}{\code{locFace},\ppvec{U}_t,\ppvec{f}^{*,\text{tmp}},\hat{\ppvec{\ell}}^+,\hat{\ppvec{\ell}}^-} $
      \EndIf
    \EndFor
    \State \Return $\ppvec{U}_t$
  \EndFunction
  \State
  \Function{DoSurfInt}{\code{locFace},$\ppvec{U}_t,\ppvec{f}^*_{pq},\hat{\ppvec{\ell}}^+,\hat{\ppvec{\ell}}^-$}
    \Switch{\code{locFace}}
      \Case{$\xi^-$}
        \For{$i,j,k \gets 0$ to $N$}
          \State $\ppvec{U}_{t,ijk} \gets \ppvec{U}_{t,ijk}+\ppvec{f}^{*}_{jk} \:\hat{\ell}_i^-$
        \EndFor
      \EndCase
      \Case{$...$}
      \EndCase
      \Case{$\zeta^+$}
        \For{$i,j,k \gets 0\text{ to }N$}
          \State $\ppvec{U}_{t,ijk} \gets \ppvec{U}_{t,ijk}+\ppvec{f}^{*}_{ij} \:\hat{\ell}^+_k$
        \EndFor
      \EndCase
    \EndSwitch
  \EndFunction
  \end{algorithmic}

 \end{algorithm}

\begin{algorithm}[h!tb]
  \caption{GPU kernel for the \code{SurfInt} operation}\label{alg:surfint_gpu}
    \begin{algorithmic}[1]
    \Kernel{SurfInt\_Kernel}{$N,n_{Elems},\ppvec{f}^{*},\ppvec{U}_t$}
      \State $i \gets$ \code{(blockID-1)*blockDim+threadID}
      \State $n_{DOF} \gets (N+1)^3 n_{Elems}$ \Comment{number of volume DOF}
      \If{$ i \leq n_{DOF}$}
\For{$\code{locFace} \in \{\xi^-,\eta^-,\zeta^-,\xi^+,\eta^+,\zeta^+\}$}
          \State $p,q,s,\hat{\ell}_k^{\pm},\code{isPrimary} \gets \Call{FaceMapping}{i,\code{locFace}}$
          \If{\code{isPrimary}}
\State $\ppvec{U}_{t,i}=\ppvec{U}_{t,i}+\ppvec{f}^{*}_{pq,s}\:\hat{\ell}_k^{\pm}$
          \Else
            \State $\ppvec{U}_{t,i}=\ppvec{U}_{t,i}-\ppvec{f}^{*}_{pq,s}\:\hat{\ell}_k^{\pm}$
          \EndIf
        \EndFor
\State \Return $\ppvec{U}_{t,i}$
      \EndIf
    \EndKernel
  \end{algorithmic}

 \end{algorithm}

In the original CPU version, \cref{alg:surfint_cpu}, the \code{SurfInt} routine loops over all faces on the current rank.
For each face, it obtains the orientation of the face with respect to the volume.
The orientation of the face of a hexahedral DG element depends on which of its six local faces it refers to.
The contribution of this face is then added to all DOFs within the element.

This operation is hard to parallelize for GPU hardware since all 6 local faces add their contribution to each individual DOF within the element.
Writing to the same entries in an array multiple times can yield race conditions if the individual threads are not properly synchronized--but synchronizing threads is costly.
In \galexi, the sequence of operations is thus altered for the GPU implementation in comparison to the original CPU implementation.
The developed algorithm, \cref{alg:surfint_gpu}, runs as follows.
First, each GPU thread is assigned a single DOF within an element.
Due to the tensor product structure of the DGSEM, this results in only a single DOF per face influencing the solution as indicated in \cref{fig:dg_cube}.
The thread then loops over all six faces (\code{locFace}) of the element.
For each face, it identifies the face index $s$ within the flux array and the corresponding DOF on the face specified by the indices $p,q$.
The face whose normal vector is used to compute the Riemann flux is determined by the flag \code{isPrimary}, while for the adjacent element (\code{isReplica} face) the sign of the flux contribution has to be flipped to account for the fact that its outward facing normal vector points in the opposite direction.
Additionally, the correct integration weight $\hat{\omega}$ is identified to add the flux contribution of this \code{locFace} to the respective DOF.
While this requires multiple threads to access the same face data multiple times, it avoids race conditions between threads without the need of explicit synchronization, since only a single thread writes to a specific entry in the $\ppvec{U}_t$ array.
Lastly, transforming the fluxes from the face-local to the element-local coordinate system requires some form of mapping.
Since \galexi is an unstructured solver, the algorithm also needs to account for the case where coordinate systems of neighboring elements are rotated with respect to each other.
The combination results in mappings which are non-trivial to obtain.
However, the required mappings are hardware-agnostic and not relevant for the efficiency of the GPU kernel.
In consequence, these specifics are condensed into a single call to a subroutine \code{FaceMapping} to keep the algorithm concise.
More details on the face connectivity can be found in \citet{krais2021flexi}.

At this point, it is important to revisit the required compute time of the different DG operations as shown in \cref{fig:galexi_piechart}.
It is evident that the majority of the computational work can be attributed to the operations \code{VolInt}, \code{Lift\_VolInt}, and \code{Cons\-To\-Prim}, which are operations local to each DG element that can be scheduled independently of any communication.
This has three crucial implications.
First, only a small number of routines require the majority of the compute time, which yields distinct targets for more sophisticated optimization.
Second, these routines do not require any communication, which again highlights the beneficial ratio of local work to required communication of DG schemes.
Third, the overhead introduced by the unstructured mesh is negligible, since the additional work is mainly limited to the routines mapping from the faces to the volumes, i.e.\ \code{SurfInt} and \code{ProlongToFace}, which take only around \SI{15}{\%} of the overall compute time.

\subsection{Summary of the Parallelization Strategy}
This section provides details on the parallelization concept of \galexi on three different levels.
First, the parallelization of the workload between GPUs was introduced.
Here, \galexi subdivides the domain into subdomains with roughly the same number of elements, which are then assigned to the individual GPUs and communication across the boundaries of neighboring subdomains is performed using CUDA-aware MPI.
Second, the individual compute kernels within the GPU are scheduled using streams to improve the overall utilization of the GPU.
Operations associated with the MPI communication are assigned to the stream with the highest priority to allow the GPU to antedate the execution of these kernels to initiate the communication at the earliest possible point in time.
Third, the design concepts of the kernels were introduced using the \code{Cons\-To\-Prim} operations as an example for pointwise operations and the \code{SurfInt} to detail the more intricate case of kernels that have to map from the elements' volume to their faces and vice versa.
The resulting performance of the kernels demonstrates that the overhead of the unstructured mesh is negligible.
A detailed discussion of the resulting parallel performance of \galexi across multiple GPUs is provided in the following paragraphs.

\section{Performance Evaluation}\label{sec:scaling}

In the following section, the performance and the scaling abilities of \galexi are demonstrated.
First, \cref{sec:hardware} introduces the details of the applied systems, i.e. HAWK-AI and JUWELS Booster.
\Cref{sec:pid} then derives the performance metrics that are used to evaluate the performance.
With these in place, \cref{sec:memory} provides details on the code's memory consumption while the results of the scaling tests are discussed in \cref{sec:scaling_results}.

\subsection{Hardware Architecture}\label{sec:hardware}
The performance of \galexi and FLEXI is investigated for two different systems.
First, the JUWELS Booster installed at the Jülich Supercomputing Centre (JSC) and second, the HAWK and HAWK-AI systems at the High-Performance Computing Center Stuttgart (HLRS).

The JUWELS Booster module entails a total of 936~two-socket nodes.
Each node provides two AMD EPYC 7402 processors with 24~cores per socket and a total of 512~GiB of DDR4-3200 main memory per node.
Each node comprises 4 NVIDIA A100 GPUs with 40 GiB memory interconeccted using NVlink, where each GPU is connected to its own network adapter and the individual nodes are integrated using a Mellanox HDR200 InfiniBand interconnect with 200 Gbit/s per adapter in a DragonFly+ topology.

The HAWK supercomputer at HLRS is based on an HPE Apollo 9000 with 5632~dual-socket nodes.
Each node is equipped with two AMD EPYC~7742 CPUs, which yield 128 CPU cores per node.
Each node comprises 256~GiB of main memory and the nodes are connected using a Mellanox HDR200 InfiniBand interconnect in a 9D-hypercube topology.
The HAWK-AI partition of HAWK is based on an HPE Apollo 6500 Gen10 Plus with 24 nodes, where each node is equipped with two 64-core AMD EPYC 7702 processors, 8 NVIDIA A100 GPUs interconeccted using NVlink, and 1~TiB of main memory.
20 nodes employ A100 GPUs with 40~GiB memory and 4 nodes entail A100 GPUs in the 80~GiB version.
The nodes of HAWK-AI are fully integrated into the main HAWK partition using an Inifiniband interconnect in a Fat-Tree topology, such that nodes from both systems can be used within a single compute job.
The HAWK-AI partition was designed to integrate AI and big data capabilities into traditional HPC jobs but is also capable of running and scaling GPU-accelerated HPC applications on its own.

\subsection{Performance Metrics}\label{sec:pid}
In the following, we focus on two distinct metrics to quantify and compare the performance of \galexi and FLEXI on different hardware, which rely on the time-to-solution and the energy-to-solution paradigms, respectively.
For this, we use the performance index (PID), which is a commonly used metric to quantify the performance of DG codes~\cite{krais2021flexi,fehn2019matrix} and is defined as
\begin{equation}
  \text{PID} = \frac{\text{Walltime}\; \times\; \#\text{Ranks}}{\#\text{RK-stages}\; \times\; \#\text{DOF}}.
\end{equation}
The PID describes the walltime required by a single rank to advance a single DOF for one stage of the explicit Runge--Kutta time-stepping.
Hence, the PID is independent of the number of timesteps performed, the number of DOF used in the simulation and the number of ranks employed, where a rank refers either to a CPU core or a whole GPU as discussed in \cref{sec:implementation}.
While this provides a good measure of efficiency for code performance comparison on either CPU or GPU systems, the usefulness of this definition is limited when comparing GPU and CPU codes with each other.
Here, a whole GPU would be compared to a single CPU core with a vastly different compute performance and power consumption.
To account for the differences in hardware, we propose in this work the novel energy-normalized PID (EPID) as a more suitable measure of performance.
The EPID is defined as\begin{equation}
  \text{EPID} = \frac{\text{Walltime}\;\times\;\text{Power}}{\#\text{RK-stages}\;\times\;\#\text{DOF}} = \underbrace{\frac{\text{Power}}{\#\text{Ranks}}}_{P_{\text{rank}}}\;\times\;\text{PID},
\end{equation}
and describes the energy required to compute the time update for a single DOF on the specific computing hardware.
The EPID can thus be interpreted as the PID normalized by the specific power required per rank, which is denoted as $P_{\text{rank}}$.

\subsection{Memory Requirements}\label{sec:memory}
The memory consumption of a real-world application on the device is given in \cref{tab:memory_requirements} in KiB per DOF for different polynomial degrees $N$.
In general, the overall memory consumption is low, which is a well-known property of the explicit numerical scheme. The results clearly show that increasing $N$ improves the memory efficiency, i.e.\ reduces the required amount of memory per DOF.
This is because \galexi stores both the solution for the DOFs within the DG element ($(N+1)^3$) and on its surfaces ($6(N+1)^2$).
With increasing $N$, the ratio between surface to volume information thus decreases, yielding a lower overall memory footprint.
As an illustration of memory efficiency, it is possible to compute a problem with $N=7$ and 48 million DOF per solution variable on a single device with 40~GiB of memory.

\begin{table*}[t]
  \centering
  \caption{Measured memory consumption per DOF on the GPU for different polynomial degrees $N$ and the Navier--Stokes equation system.}\label{tab:memory_requirements}
  \begin{tabular}{ l c c c c c c c c c c c c }
    \toprule
    N   & 1 & 2 & 3 & 4 & 5 & 6 & 7 & 8 & 9 & 10 & 11 & 12 \\
    \midrule
    KiB & 1.457 & 1.188 & 1.049 & 0.996 & 0.942 & 0.895 & 0.869 & 0.841 & 0.827 & 0.808 & 0.801 & 0.787 \\
    \bottomrule
  \end{tabular}
\end{table*}

\subsection{Scaling Tests}\label{sec:scaling_results}

To evaluate the scalability of \galexi on HPC systems, its parallel performance is evaluated on the JUWELS booster module using up to 1024 GPUs for a wide range of problem sizes.
For this, the spatial resolution of a Cartesian mesh with $4\times4\times2=32$ elements is successively doubled in each spatial direction until the finest resolution of $256^3=\num[round-mode=places,round-precision=1]{16.777216e6}$ elements is reached.
For a polynomial degree of $N=7$, which is a typical choice for production runs, this results in \num{16384} to \num[round-mode=places,round-precision=1]{8.589934592e9} DOF, respectively.
The case for the scaling test is based on the setup of the incompressible Taylor--Green-Vortex considered in \cref{sec:validation:tgv}. Its simulation domain is a triple periodic box with corresponding initial conditions. However, in contrast to an implicit time integration scheme, the computational cost and thus the scaling behavior of the explicit scheme is independent of the chosen initial condition.
Each computation is advanced for 100 timesteps and the scaling properties are evaluated based on the PID.
Here, only the time for the timestepping is considered and initialization and analyze routines are neglected.
The results of the scaling tests are presented from three different perspectives---first, the influence of the computational load per GPU on the overall performance, second, investigating the parallel efficiency in a weak scaling setting and third, from a strong scaling perspective.

\begin{figure}[t]
  \centering
  \includegraphics{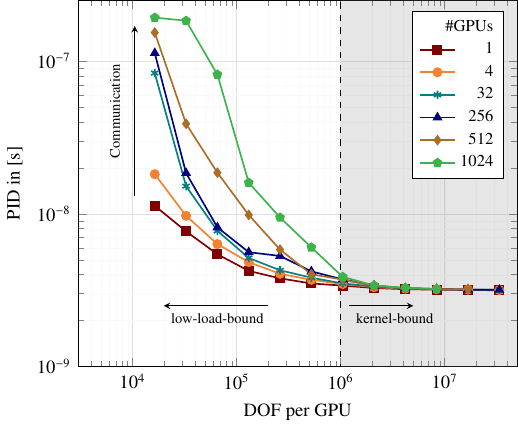}   \caption{Scaling results for \galexi with the split-form DG scheme and $N=7$ plotted as PID over the specific load, i.e.\ DOF per GPU, for up to 1024 GPUs.}\label{fig:scaling_PID}
\end{figure}
In a first step, the PID is plotted against the specific load in terms of DOF per GPU in \cref{fig:scaling_PID}.
Since the PID is a measure of computational time, a lower PID indicates better performance. Most strikingly, all curves converge above the limit of \num{e6} DOF per GPU, which means that the overhead of the parallelization and communication becomes negligible in comparison to using only a single GPU.
Hence, \galexi scales almost perfectly beyond the threshold of \num{e6} DOF per GPU.
The behavior changes for loads below this threshold.
Here the PID increases towards lower loads for all cases, which means that the computational efficiency decreases.
Moreover, the more GPUs are used for the simulation, the more pronounced this loss in performance becomes.
This can be attributed to two factors.
First, the communication latency between the GPUs cannot be hidden completely at low loads, since the amount of local work is insufficient to hide the communication.
Furthermore, the loss in performance becomes more pronounced the more potential communication partners, i.e. GPUs, are used for the simulation.
The severity of this performance penalty depends strongly on the network topology of the HPC system and the job placement on the system, which is determined by the scheduler.
In the case of the JUWELS booster module, which uses a DragonFly-type network topology, the communication cost increases significantly when the nodes are spread across a larger number of switch groups, which contain 192 GPUs each.
However, lacking latency hiding cannot explain the performance loss when using a single GPU, since here no communication is necessary.
Instead, this drop in performance can be attributed to the overhead associated with launching kernels on the GPU.
If the actual computational load of the kernel becomes too small, the kernels cannot be launched quickly enough to use the GPU to capacity.
Moreover, tail effects become noticeable, as discussed in \cref{sec:implementation}.
To summarize the results, the GPU implementation can be seen to be \emph{kernel-bound} for high loads, where the performance becomes independent of the total number of GPUs used. For very low loads, the performance gets \emph{low-load-bound} and becomes increasingly \emph{communication-bound} with a dominant performance penalty the more compute nodes are used.
This is in stark contrast to the CPU implementation of FLEXI as reported by \citet{blind2023towards}.
Here, the impact of the communication overhead is similarly noticeable for very low loads.
However, a performance penalty also appears for very high loads, since here the fast CPU cache cannot hold all necessary data and the bandwidth to the main memory becomes the bottleneck.
This results in a narrow band in the rage of \num{3000} to \num{10000} DOF per rank, where optimal performance is achieved \cite{krais2021flexi,blind2023towards}. In the case of \galexi, increasing the load only improves the overall performance with the available GPU memory as the single limiting factor.

\begin{figure}[t]
  \centering
  \includegraphics{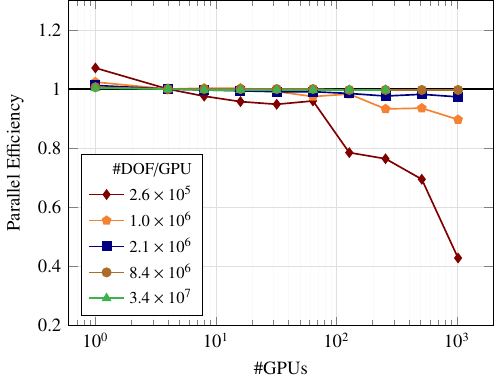}

   \caption{Weak scaling of \galexi with the split-form DG scheme and $N=7$ plotted as the parallel efficiency over the number of GPUs for specific loads, i.e. DOF per GPU. The parallel efficiency is computed based on the PID on a single node, i.e.\ on 4 GPUs.}\label{fig:scaling_weak}
\end{figure}

In \cref{fig:scaling_weak} the investigated weak scaling properties of \galexi are depicted.
In the weak scaling paradigm, the problem size and the amount of compute resources are increased proportionally, such that the overall load per GPU is kept constant for each case.
The parallel efficiency is presented with respect to the performance of a complete compute node equipped with 4 GPUs (parallel efficiency equal to 1). This is done in order to take communication into account in a meaningful way, as the baseline case includes MPI communication.
It also enables a suitable assessment of the communication overhead in the case that only one GPU is used without communication. The results show again the threshold of \num{e6} DOF per GPU as discussed before.
For lower loads, the communication latency degrades the overall performance, while loads above \num{e6} DOF per GPU show almost perfect weak scaling up to the maximum of \num{1024} GPUs.

\begin{figure}[t]
  \centering
  \includegraphics{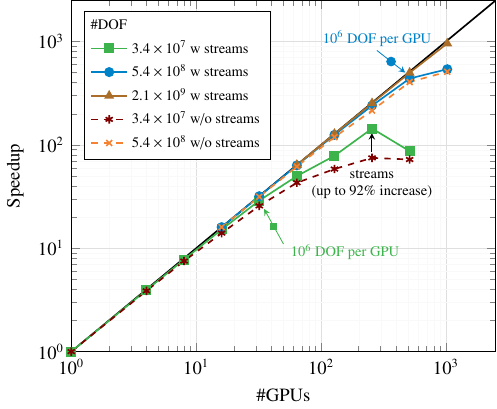}   \caption{Strong scaling of \galexi with the split-form DG scheme and $N=7$ plotted as the speedup over the number of GPUs for three problem sizes. For two cases, the results without the use of parallel streams are shown dashed. The speedup is computed based on the smallest number of GPUs that was able to run the given case. The ideal speedup is shown in black.}\label{fig:scaling_strong}
\end{figure}

Lastly, the results for strong scaling of \galexi are shown in \cref{fig:scaling_strong}.
The strong scaling analysis investigates how performance changes when the number of GPUs is increased while the problem size remains constant.
This causes problems for large cases, which might not fit into the memory of only a few GPUs, but rather require a larger number of GPUs to be actually computed.
This limitation is the reason why the scaling results in \cref{fig:scaling_strong} not exclusively start at a single GPU, but rather at the minimum number of GPUs required to compute the given problem size due to memory constraints.
The performance using this minimum number of GPUs is then used to compute the relative speedup when increasing the number of GPUs.
The strong scaling capabilities of \galexi are excellent up to the maximum 1024 GPUs, as long as the computational load exceeds the threshold of \num{e6} DOF per rank, which is indicated explicitly for both cases.
Below this threshold, i.e.\ towards larger number of GPUs, the load per device is insufficient to exploit the computing power of the GPU and to hide the necessary communication, which results in the loss of performance.
This also matches the results by \citet{fischer2022nekrs}, who report that NekRS reaches its limit for strong scaling at a similar load of about 2 to 4 million DOF per rank.
For computational loads above this threshold, \galexi yields almost perfect strong scaling results up to the maximum of 1024 GPUs.
Next, the influence of our scheduling strategy based on parallel streams and introduced in \cref{sec:implementation_gpu} is investigated.
For this, \cref{fig:scaling_strong} also shows the scaling results for the same problem sizes, with (solid lines) and without (dashed lines) the use of parallel streams for kernel scheduling.
For both setups, the omittance of stream scheduling results in a significant loss in parallel performance for low loads.
This can be attributed to two aspects.
First, parallel streams allow for hiding the overhead of kernels launches and tail effects for low loads.
However, more importantly, our implementation permits the GPU to preempt the computation of quantities that have to be communicated via MPI.
This facilitates more efficient communication latency hiding, resulting in better parallel performance in cases involving many communication partners and low amounts of local work.

\section{Verification \& Validation}\label{sec:validation}

\subsection{Verification - Convergence Tests}\label{sec:validation:convergence}

\begin{figure*}[t]
  \centering
  \includegraphics{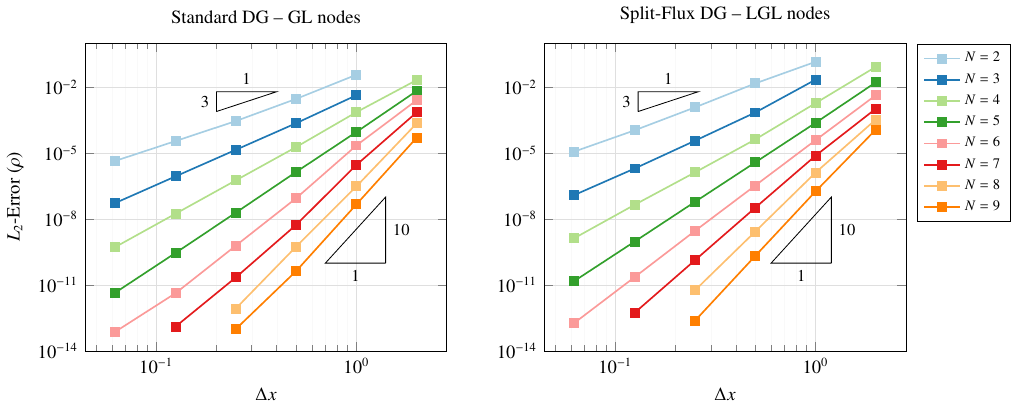}   \caption{Convergence of the split-flux DG scheme on LGL nodes (\emph{left}) and the standard DG scheme on GL nodes (\emph{right}) using $N\in[2,9]$ for the manufactured solution.}\label{fig:convtest}
\end{figure*}

The correct implementation of the high-order accurate numerical schemes in \galexi is verified by testing the order of convergence of the spatial operator with the method of manufactured solutions~\cite{roache2002code}.
This method allows the derivation of source terms for nonlinear partial differential equations that lead to exact solutions that can be expressed in analytical form and allow computing the error of the numerical discretization scheme.
Following \citet{hindenlang2012}, the exact function is assumed to follow a sinusoidal solution of the form
\begin{align}
       \rho(\ppvec{x},t) &= \phantom{\big{(}} 2+A\sin\left(2\pi(x+y+z-at)\right)        ,\nonumber       \\
  \ppvec{u}(\ppvec{x},t) &= \phantom{\big{(}} 2+A\sin\left(2\pi(x+y+z-at)\right)        , \label{eq:mms} \\
          E(\ppvec{x},t) &=          \big(    2+A\sin\left(2\pi(x+y+z-at)\right)\big)^2 ,\nonumber
\end{align}
where the amplitude and advection speed are chosen as $A=0.1$ and $a=1$, respectively.
This solution describes an oblique, periodic wave that is advected linearly with speed $a$.
The source terms that are required for \cref{eq:mms} to be an exact function of the NSE are detailed in \citet{gassner2009polymorphic}.
The problem is then initialized within a domain of $\ppvec{\Omega}\in[-1,1]^3$ with periodic boundary conditions and is discretized with varying $N\in[2,9]$.
The meshes are varied in the range of containing a single element up to $64^3$ elements at maximum.
The computation is advanced in time up to $t=1$ and the timestep is chosen sufficiently small to not influence the overall discretization error.
The convergence test is carried out with both the standard collocation formulation on GL interpolation points and the split-flux formulation on LGL interpolation points.
The results in \cref{fig:convtest} demonstrate that the expected design order is reached for all investigated cases, which verifies the correct implementation of the schemes.

\subsection{Validation - Taylor--Green-Vortex}\label{sec:validation:tgv}

A popular validation case for turbulent flows is the Taylor--Green-Vortex~(TGV) introduced by Taylor and Green \cite{taylor1937mechanism}.
One reason for its widespread use are its analytically prescribed initial conditions, which are given for a domain of size $\ppvec{\Omega}\in[0,2\pi L]^3$ by
\begin{align}
  \label{eq:tgv_ic}
  \ppvec{u}(\ppvec{x},0) &=
    \left(
    \begin{matrix}
      \phantom{-}U_0 \sin\left(\frac{x}{L}\right)\cos\left(\frac{y}{L}\right)\cos\left(\frac{z}{L}\right)\\
                -U_0 \cos\left(\frac{x}{L}\right)\sin\left(\frac{y}{L}\right)\cos\left(\frac{z}{L}\right)\\
                 0
    \end{matrix}
    \right),\\
  p(\ppvec{x},0) &= p_0 + \frac{\rho_0 U_0^2}{16} \left(\cos\left(\tfrac{2x}{L}\right)+\cos\left(\tfrac{2y}{L}\right)\right)\left(2+\cos\left(\tfrac{2z}{L}\right)\right)\nonumber,\end{align}
with $L=1$ denoting the characteristic length, $U_0=1$ the magnitude of the initial velocity fluctuations and $\rho_0=1$ the reference density.
The background pressure $p_0$ is chosen to fit a prescribed background Mach number $\ppMa_0=U_0\sqrt{\rho_0/(\gamma p_0)}$ and the viscosity $\mu_{ref}$ is used to obtain the desired Reynolds number which is defined as $\ppRe=\rho_0 U_0 L / \mu_{ref}$. However, \cref{eq:tgv_ic} does not yield sufficient initial conditions for a compressible flow field, since it lacks information about the density and temperature fields.
Two different approaches are commonly used to extend it to a full description of a compressible flow field as required for the computation with a compressible solver.
For this, either field is held constant, while the other quantity is computed to yield a thermodynamically admissible state.
Assuming an perfect gas that follows \cref{eq:eos_T} this yields the two variants
\begin{align}
  \text{Version \makebox[\widthof{II}][r]{I}:}  \quad & \rho(\ppvec{x},0) = \rho_0,         & & T(\ppvec{x},0)=\frac{p}{R\rho_0} ,\\
  \text{Version II:}                            \quad & \rho(\ppvec{x},0) = \frac{p}{RT_0}, & & T(\ppvec{x},0)=T_0. \label{eq:version2}
\end{align}

Two common metrics to assess the accuracy of numerical schemes for the TGV case are the instantaneous kinetic energy in the domain $E_k$ and the viscous dissipation rate $\varepsilon_T$. The integral kinetic energy is defined as
\begin{equation}
  E_k = \frac{1}{2 \rho_{0} U_0^2\left|\Omega\right|} \int_\Omega \rho \, \ppvec{u} \cdot \ppvec{u} \,\mathrm{d}\Omega,
\end{equation}
where $\left|\Omega\right|$ denotes the overall size of the integration domain.
The viscous dissipation rate of the kinetic energy can be split into a solenoidal and a dilatational contribution (Zeman~\cite{zeman1990dilatation}, Sarkar et al.~\cite{sarkar1991analysis}), which are defined as
\begin{align}
  \varepsilon_S &= \frac{L^2}{ \ppRe U_0^2 \left|\Omega\right|} \int_\Omega \frac{\mu(T)}{\mu_0} \,\ppvec{\omega} \cdot \ppvec{\omega}\,\mathrm{d}\Omega ,\\
  \varepsilon_D &= \frac{4L^2}{3 \ppRe U_0^2 \left|\Omega\right|} \int_\Omega \frac{\mu(T)}{\mu_0} \,\left(\nabla \cdot \ppvec{u}\right)^2\mathrm{d}\Omega ,
\end{align}
respectively.
The solenoidal component $\varepsilon_S$ can be related to the vortical motion and the dilatational component $\varepsilon_D$ to compressibility effects.

Two versions of the TGV case are investigated, which both exhibit a Reynolds number of \mbox{$\ppRe=\num{1600}$} with the initial conditions prescribed in \cref{eq:tgv_ic}.
First, the weakly compressible case with $\ppMa_0=0.1$ is investigated to verify that \galexi accurately captures the physics of turbulent flow.
In a second step, the Mach number is increased to $\ppMa_0=1.25$, which causes complex shock patterns to emerge during the simulation.
Consequently, this supersonic TGV setup is a suitable test case to assess the stability and accuracy of compressible flow solvers for shock-turbulence interaction.

\begin{figure}[t]
	\centering
  \resizebox{\linewidth}{!}{\includegraphics{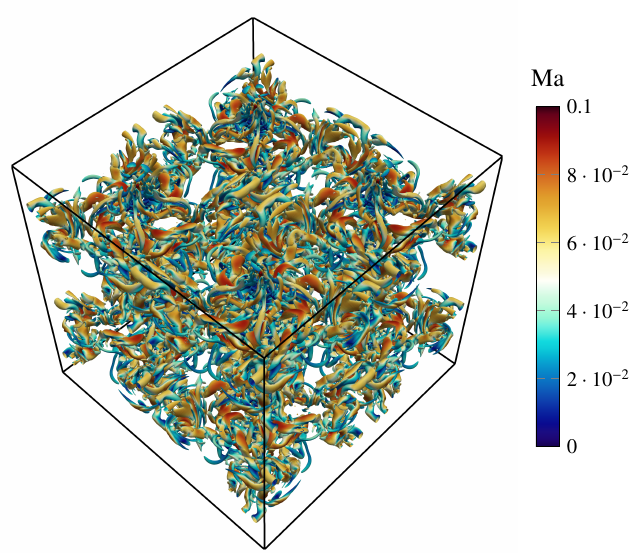}
   }
  \caption{Instantaneous flow field of the incompressible TGV case with $\ppMa_0=0.1$ and $\ppRe=1600$ at time $t=10$ using $512^3$ DOF visualized by iso-surfaces of the Q-criterion colored by $\ppMa$.}\label{fig:tgv_field}
\end{figure}

\begin{figure*}[t]
  \centering
  \includegraphics{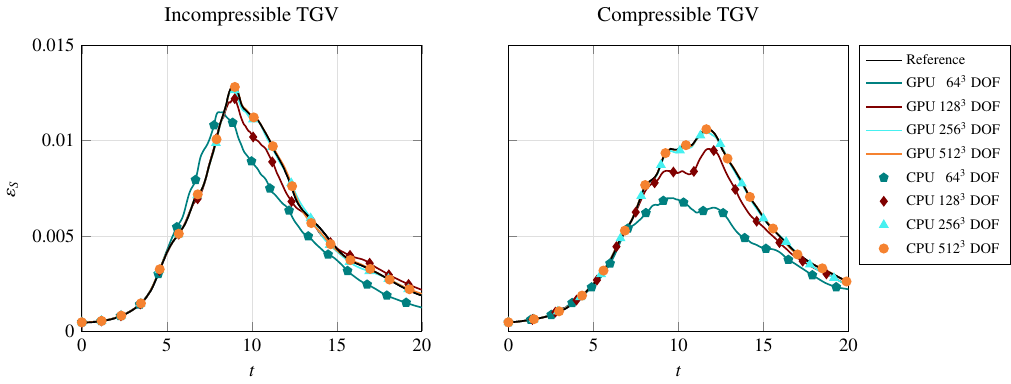}

   \caption{Temporal evolution of the solenoidal dissipation rate $\varepsilon_S$ for the incompressible TGV at $\ppMa_0=0.1$ (\emph{left}) and the compressible TGV at $\ppMa_0=1.25$ (\emph{right}) using between $8^3$ and $64^3$ elements with a polynomial degree $N=7$ which results in $64^3$ to $512^3$ DOF. The results by \citet{debonis2013solutions} (\emph{left}) and \citet{chapelier2024TGV} (\emph{left}) serve as the reference solution for the incompressible and compressible case, respectively. The results of the CPU implementation are given for reference.}\label{fig:tgv}
\end{figure*}

\subsubsection*{Incompressible TGV}\label{sec:tgv_incomp}

First, we consider the TGV at $\ppRe=1600$ in the incompressible limit with $\ppMa_0=0.1$ and Version II, i.e.\ an initially constant temperature field.
Four different uniform resolutions were investigated to demonstrate the mesh convergence of the code.
For this, either $8^3$, $16^3$, $32^3$, or $64^3$ elements were used with a polynomial degree of $N=7$, corresponding to $64^3$, $128^3$, $256^3$, or $512^3$ DOF.
Two simulations were carried out for each mesh, first with the GPU-accelerated \galexi and second with its CPU-based predecessor FLEXI for verification purposes.
The results are also validated against the high-fidelity reference solution published by DeBonis~\cite{debonis2013solutions}.
The results shown in \cref{fig:tgv} (\emph{left}) demonstrate that \galexi and FLEXI yield the same results up to machine precision.
Moreover, as the resolution increases, the temporal evolution of the dissipation rate converges to the reference solution, to the point where the solution on the finest mesh with 512 DOF in each direction matches the reference almost perfectly.
The instantaneous flow field of the TGV case at $t=10$ is illustrated in \cref{fig:tgv_field}, highlighting its vortex structures using iso-contours of the Q-criterion colored by the Mach number.
\subsubsection*{Compressible TGV}\label{sec:tgv_comp}

More recently, the TGV case was extended to the compressible regime by increasing the Mach number of the initial flow field~\cite{lusher2021assessment, chapelier2024TGV}.
A common choice is $\ppMa_0=1.25$, for which complex shock patterns emerge that interact with the turbulent flow.
Consequently, the compressible, supersonic TGV case allows for assessing the stability and accuracy of compressible flow solvers for shock-turbulence interactions.
The simulation is again initialized using the setup in \cref{eq:version2}, i.e.\ Version II, and Sutherland's law is applied to address the dependency of the viscosity on the temperature in the compressible case.
The shock capturing scheme introduced in \cref{sec:shock_capturing} is applied for the stabilization of the scheme near shocks.
Again, four mesh resolutions were investigated with $64^3$, $128^3$, $256^3$, and $512^3$ DOF and a polynomial degree of $N=7$.
The permitted maximum of the blending parameter $\alpha$ is set identically across all investigated resolutions.
The results reported by \citet{chapelier2024TGV} serve as the reference solution.
The results in \cref{fig:tgv} (\emph{right}) again show that \galexi and FLEXI yield identical results for the temporal evolution of the solenoidal dissipation rate.
Moreover, at higher resolutions, the results converge to the reference solution, where the results are almost identical for the largest case of 512 DOF per spatial direction.

\section{Application}\label{sec:application}

Based on these verification and validation results, both \galexi and FLEXI are applied to the large-scale application case of a wall-resolved LES of the NASA Rotor \num{37}~\cite{Reid1978}.
This allows for verification that \galexi can handle complex simulations of compressible flow and quantify the gains in efficiency and energy-to-solution by using GPUs.
For this, \cref{sec:app_description} first provides some background on the case, while \cref{sec:app_setup} gives details on the computational setup.
Finally, the results are discussed in \cref{sec:app_results}.

\subsection{Description}\label{sec:app_description}
In the following section, the applicability of \galexi towards large-scale test cases is demonstrated for the turbulent flow within a NASA Rotor \num{37} rectilinear transonic compressor cascade.
This rotor was originally employed in one of four transonic axial-flow compressor stages designed and tested at the NASA Lewis Research Center in the late 1970s~\cite{Reid1978}.
With its geometry parameters and measurement data publicly available~\cite{Moore1980,Suder1996}, the rotor has since become a benchmark test case in the turbomachinery research community including CFD studies~\cite{Denton1997}, investigation of optimization techniques~\cite{Benini2004}, tip leakage flow analysis~\cite{Seshadri2015}, and uncertainty quantification approaches~\cite{Loeven2010}.
At its design point, the rotor operates with a blade tip Mach number of \num{1.4939}, generating an overall pressure ratio of \num{2.106}.
The setup investigated here corresponds to a ground-idle condition, providing a tip Mach number of \num{0.824} with a total pressure ratio of \num{1.305}.
The cascade geometry is generated by unwinding the blade profile at mid-span and extruding for \SI{5}{\percent} of the chord length.
The resulting Reynolds number based on the inflow velocity and the rotor chord is \num{972550}.
The low operating point and the position at mid-span results in an inlet relative Mach number of \num{0.758} and an incidence relative to the mean camberline of \SI{10.1}{\degree}.
Validation of the FLEXI solution against experimental integral data at mid-span is given in \citet{Kopper2024}.
Results obtained from \galexi were confirmed to match the solution of the CPU-based framework.
Both solvers predict a transonic expansion region forms on the suction side near the leading edge as a result of the high subsonic inflow velocity and near-stall condition.
The region is terminated with a near-normal shock and subsequent shock-boundary-layer interaction with flow separation occurring throughout the suction side.
On the pressure side, a small laminar separation region forms which is subsequently terminated by turbulent re-attachment.

\subsection{Computational Setup}\label{sec:app_setup}

\begin{figure}[t]
  \centering
	\includegraphics{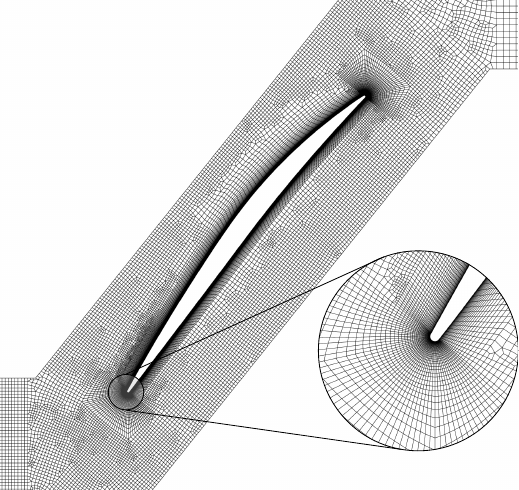}
 \caption{Computational mesh for the simulation of the NASA Rotor 37 case. The inflow and outflow regions are pruned and a zoom highlights the mesh around the leading edge.}\label{fig:nasa_mesh}
\end{figure}

The computational setup is identical for both \galexi and FLEXI, except for the hardware on which the simulations are run.
The mesh for the LES comprises one compressor pitch with the compressor blade orientated with the stagger angle of \SI{51.2}{\degree} and is depicted in \cref{fig:nasa_mesh}.
The domain is discretized using \num[round-mode=places,round-precision=1]{1.202760e6} elements with $N=5$, which results in a total of \SI[round-mode=places,round-precision=1]{2.59796160e8}{DOFs} for the simulation.
The inflow is modeled using far-field conditions and a subsonic outflow condition~\cite{carlson2011inflow} is employed.
Additionally, sponge zones~\cite{flad2014discontinuous} are positioned at the inflow and outflow boundaries to prevent the formation of artificial reflections.
The rotor itself is modeled as an adiabatic wall and the spanwise and pitchwise boundaries are defined as periodic. The simulation is performed using the split-form DG method as introduced in \cref{sec:dgsem} to mitigate aliasing errors with the flux formulation given by \citet{pirozzoli2011numerical}.
The solution is advanced in time using a 14-stage \nth{4}-order Runge--Kutta method~\cite{niegemann2012efficient}.
During the simulation, the viscosity is computed with Sutherland's law as given in \cref{eq:sutherland}.

\begin{figure*}[t]
	\centering
  \resizebox{\textwidth}{!}{\includegraphics{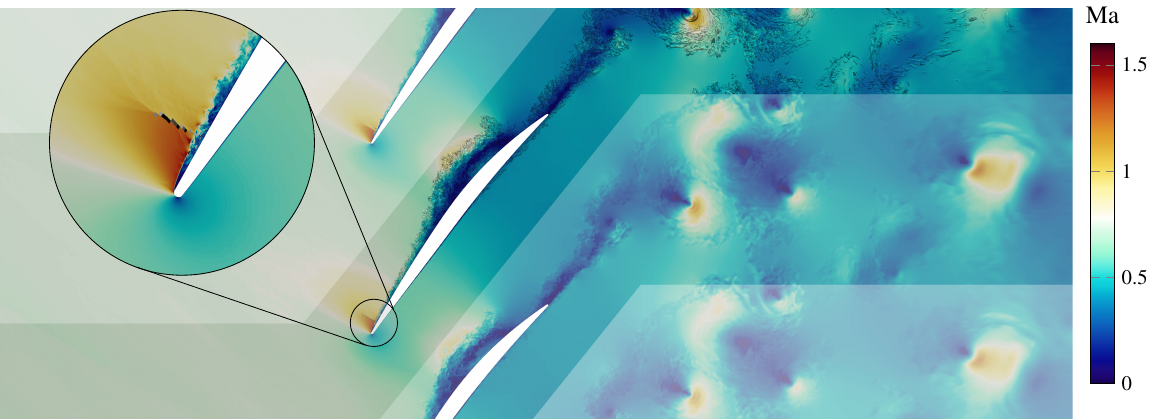}
 }
  \caption{Instantaneous field solution for the NASA Rotor 37 case colored by the Mach number. A zoom of the leading edge highlights the supersonic flow region with the local blending values $\alpha$ of the FV shock capturing scheme overlaid for all elements with $\alpha=0.1$ (light gray) up to $\alpha=0.7$ (black). The domain is periodically extended which is indicated by a blurred overlay.}\label{fig:nasa_sol}
\end{figure*}

\begin{table*}[t]
\centering
  \caption{Setup and performance results for the simulation runs on both CPU and GPU for a simulation time of $8t^\ast$.}
  \begin{tabular}{lrcccccccc}
    \toprule
        &       Ranks &                   DOF/Rank & $P_{\!\text{rank}}$ [W] &                     PID [s] &                  EPID [J] & Walltime/$t^\ast$ [s] & Energy/$t^\ast$ [kWh] \\ \midrule
    GPU & \num{  128} & \numTwoPlaces{2.0296575e6} &              \num{448}  & \numTwoPlaces{4.579389e-09} & \numTwoPlaces{2.0505e-06} &         \num{9209}    &          \num{147} \\ CPU & \num{32768} & \numTwoPlaces{7.92835e3}   &   \numTwoPlaces{4.9414} & \numTwoPlaces{1.024092e-06} & \numTwoPlaces{5.0604e-06} &         \num{7538}    &          \num{339} \\ \midrule
Savings &         &                            &                         &                             &             \SI{59.5}{\%} &                       &      \SI{56.8}{\%} \\
    \bottomrule
  \end{tabular}\label{tab:app:cpu_gpu}
\end{table*}

The computational resources are chosen such that both codes run at their maximum efficiency.
For \galexi, \num{128} Nvidia A100 GPUs on HAWK-AI are employed, which yields a total load of \SI[round-mode=places,round-precision=1]{2.029658e6}{DOF} per GPU.
For FLEXI, the number of CPU nodes is chosen such that the walltime is similar to the \galexi computation, which is obtained when using \num{256} nodes (\num{32768} CPU cores).
This results in a load of around \SI{7900}{DOF/core}, which resides well within the performance optimum of FLEXI~\cite{krais2021flexi}.
The details of these setups are summarized in \cref{tab:app:cpu_gpu}.
The simulations are initialized with a precomputed converged flow state and are advanced for a total of 8 characteristic time units~$t^\ast=t\,u_{\infty}/c$, where $t^\ast$ is defined with respect to the inflow velocity $u_{\infty}$ and the chord length $c$.

\subsection{Results}\label{sec:app_results}

The instantaneous Mach number distribution on the domain centerline computed by \galexi is depicted in \cref{fig:nasa_sol}.
The flow enters from the left with a high incidence relative to the camberline.
This results in a shift of the stagnation point towards the pressure side and a strong transonic expansion fan on the suction side.
The supersonic region is terminated with a near-normal shock, as illustrated in the zoom region.
The corresponding pressure jump results in a forced boundary layer transition with high levels of unsteadiness.
Numerical oscillations in the vicinity of the discontinuity, i.e.\ shock, resulting from Gibb's phenomenon, are mitigated with the convex blending approach outlined in \cref{sec:shock_capturing}.
The grayscale overlay in the zoom region represents the local values of the blending factor $\alpha$.
It is evident that the FV shock capturing is active only near the shock in order to preserve the high numerical order of the DG operator in areas with a smooth solution.
Downstream of the shock region, the separation of the boundary layer causes temporally varying blockage which couples with the upstream flow physics resulting in a highly unsteady flow field.
Periods with enhanced separation result in counter-rotating vortex shedding as is visible near the wake downstream of the blade row.

The achieved performance for both codes is summarized in \cref{tab:app:cpu_gpu}.
For \galexi, a PID increase of about \SI{35}{\percent} is observed in comparison to the performance reported in \cref{sec:scaling}.
This is attributed to the additional work and load imbalance between the ranks introduced by the test case, which includes the sponge zones, the boundary conditions, the shock indicator and the FV shock capturing scheme. The slight deviation in walltime per $t^*$ between the GPU and CPU cases stems from choosing powers of two for the resources.

The power draw of the simulation runs was measured via the monitoring facilities of the systems' operators at the HLRS.
It is important to stress that these measurement systems are not originally designed for the purpose of measuring the energy consumption of individual simulation runs, but rather to monitor and adjust the power draw of the overall facility.
As a consequence, the power draw can only be measured on a rack-wise level and is limited in terms of accuracy and granularity, which means that the obtained results should be seen as a rough estimate.
From these measurements, the specific power draw per rank $P_{\!\text{rank}}$ shown in \cref{tab:app:cpu_gpu} is computed as the overall power delivered to the racks used divided by the number of ranks
The measured power thus also includes the power for the network switches.
It is important to note that due to the specific hardware layout, cooling is included in the total power consumption for the GPU case, while the cooling effort is not included for the CPU system.
Hence, the obtained results tend to favor the CPU implementation and should thus be seen as a conservative lower bound for the potential gains in efficiency provided by GPU hardware.

When comparing the resulting EPID, i.e.\ the necessary amount of energy to advance a single DOF for a single time level, \galexi more than halves the required energy-to-solution.
In total, \galexi requires around \SI{147}{kWh} to advance the solution for one characteristic time unit $t^\ast$, while FLEXI requires around \SI{339}{kWh} per $t^\ast$ on CPUs.
It is reasonable to relate this reduction in energy demand by \galexi to a similar reduction in associated carbon emissions.
However, it is important to note that the I/O operations and analyzing routines are excluded from the PID computation.
Since these operations are still performed on the CPU for \galexi, the resulting overhead causes a slight discrepancy in the savings for the EPID and energy-to-solution.
As discussed before, due to the measurement limitations, both results should be regarded as an estimate and lower bound of the achieved performance.

\section{Conclusion \& Outlook}\label{sec:conclusion}

This work presents the open-source flow solver \galexi, which implements high-order DG methods on unstructured mesh\-es for GPU-accelerated HPC systems.
\galexi is the GPU-accelerated spinoff of the established FLEXI solver and it supports the majority of the features provided by FLEXI, which are continuously being extended.
This allows the application of \galexi for scale-resolving simulations of complex compressible flows including shock waves using modern GPU-based HPC systems.
This work provides details on the general code design, the parallelization strategy, and the implementation approach for the compute kernels.
Thus, it serves as an indication on how existing spectral element codes can be ported efficiently for GPUs.
As long as the GPUs are sufficiently loaded, the results demonstrate excellent scaling properties for \galexi on up to 1024 GPUs.
The correct high-order accurate implementation of \galexi has been verified by demonstrating the expected convergence rates.
Furthermore, the code has been validated against reference data for the incompressible and compressible variants of the established TGV.
As a demonstration of a large-scale application, \galexi was employed for the simulation of a wall-resolved LES of a NASA Rotor 37 compressor cascade.
Using this example of compressible flow, the implemented finite volume subcell approach was demonstrated to yield a stable and accurate scheme for capturing the unsteady supersonic expansion region at the leading edge.
In addition, \galexi has been shown to use only half the energy required to run the same simulation using the CPU implementation.
With this, \galexi reduced the required energy from around \SI{339}{kWh} to \SI{147}{kWh} per characteristic time unit in comparison to the CPU implementation, which halved the associated carbon emissions.

Currently, \galexi is implemented using the CUDA Fortran framework, which does not support GPU hardware from vendors other than NVIDIA.
Current efforts are focused on incorporating different compute backends into \galexi to support accelerator devices of different vendors alongside the baseline CPU implementation via hardware abstractions.
The envisioned code is intended to be readily extendable to arbitrary compute devices, such that novel accelerator types can be incorporated without fundamental code redesigns.
Concurrent work focuses on further optimization of key routines, in particular the \code{VolInt} and \code{FillFlux} routines, which together consume almost half of the computing time as was demonstrated. Along the same lines, automatic tuning of hardware-specific launch configurations is to be integrated into the code.
This is expected to provide high levels of performance across a wide range of different hardware.
Here, the KernelTuner~\cite{kerneltuner} package appears to be a suitable choice.
Lastly, graph-based approaches to domain decomposition might improve the utilization of the direct, high-bandwidth connection between individual GPUs on the same node by maximizing the amount of intra-node and minimizing the amount of inter-node communication.

This work has demonstrated that high-order DG methods are well-suited candidates for the efficient simulation of compressible flows on GPU systems.
\galexi has showcased that unstructured mesh topologies and adequate state-of-the-art shock capturing based on FV subcells impose only negligible overhead on GPU hardware.
Most importantly, \galexi is capable of reducing the carbon emission associated with large-scale flow simulations by more than \SI{55}{\%} in comparison to the CPU reference, which renders it a potent tool for the upcoming generation of sustainable, exascale HPC systems.

\section*{Acknowledgments}

This work was funded by the European Union.
This work has received funding from the European High Performance Computing Joint Undertaking (JU) and Sweden, Germany, Spain, Greece, and Denmark under grant agreement No 101093393.
Moreover, this research presented was funded by Deutsche For\-schungs\-ge\-mein\-schaft (DFG, German Research Foundation) under Germany's Excellence Strategy EXC 2075 -- 390740016, by the DFG Rebound -- 420603919, and in the framework of the research unit FOR 2895.
We acknowledge the support by the Stuttgart Center for Simulation Science (SimTech).
The authors gratefully acknowledge the Gauss Centre for Supercomputing e.V. (www.gauss-centre.eu) for funding this project by providing computing time through the John von Neumann Institute for Computing (NIC) on the GCS Supercomputer JUWELS~\cite{JUWELS} at Jülich Supercomputing Centre (JSC) as well as the support and the computing time on ``Hawk'' and its ``Hawk-AI'' extension provided by the Supercomputing Centre Stuttgart (HLRS) through the project ``hpcdg''.
This work was completed in part at the Helmholtz GPU Hackathon, part of the Open Hackathons program. The authors would like to acknowledge \href{https://openacc-standard.org}{OpenACC-Standard.org}, JSC, HZDR, and HIDA for their support.

\section*{Data Availability Statement}
The \galexi and FLEXI codes used within this work are available under the \href{https://www.gnu.org/licenses/gpl-3.0.html}{GPL-3.0} license at:
\begin{itemize}
  \item \url{https://github.com/flexi-framework/galaexi}
  \item \url{https://github.com/flexi-framework/flexi}
\end{itemize}

The data generated in the context of this work and instructions to reproduce them with these codes are made available under the \href{https://creativecommons.org/licenses/by/4.0/}{CC-BY~4.0} license sorted by section at:
\begin{itemize}
  \item \href{https://doi.org/10.18419/darus-4140}{{\tt 10.18419/darus-4140}} (\cref{sec:scaling})
  \item \href{https://doi.org/10.18419/darus-4155}{{\tt 10.18419/darus-4155}} (\cref{sec:validation:convergence})
  \item \href{https://doi.org/10.18419/darus-4139}{{\tt 10.18419/darus-4139}} (\cref{sec:validation:tgv})
  \item \href{https://doi.org/10.18419/darus-4138}{{\tt 10.18419/darus-4138}} (\cref{sec:application})
\end{itemize}

\appendix
\section{Description of the Routines for the Discontinuous Galerkin Spectral Element Method}\label{app:dg_routines}

The following paragraph provides more detailed descriptions of the individual routines required to evaluate the three-dimensional DG operator given in \cref{eq:dgsem}.
Please note that only the main routines listed in \cref{tab:dg_routines} are detailed here, which are the both the most important routines and also the ones discussed throughout the paper.
Consequently, the lifting operator is omitted in the following.
The routines are listed in the order of their computation in the DG scheme.

\subsubsection*{\code{ConsToPrim}}
This routine computes the primitive variables $\ppvec{U}^{prim}$ from the state $\ppvec{U}$.
The primitive variables are required for two different purposes.
First, the computation of the pressure $p$ from the conserved variables $\ppvec{U}$ is required to evaluate the fluxes of the NSE.
Secondly, the primitive variables are needed for the computation of the gradients in the lifting routines, which are required to evaluate the viscous fluxes.
For an ideal gas, the primitive variables can be computed using the EOS in \cref{eq:eos_p,eq:eos_T}.
The computation of the primitive variables is DOF-local, since it only depends on the conserved state $\mathbf{U}$ at a given point and can thus be performed independently for each interpolation point in the volume and on the element faces.

\subsubsection*{\code{VolInt}}
The \code{VolInt} is the computationally most intense routine in the DGSEM.
It can be separated into two parts.
First, the evaluation of the fluxes $\ppvec{\mathcal{F}}^{1,2,3}$ at each interpolation point of the element and second, the linewise multiplication of the fluxes with the derivative matrix $\hat{\ppvec{D}}$, which entails the derivatives of the basis functions.
The evaluation of the fluxes is again DOF-local and can be performed independently for each interpolation point.
In contrast, the application of the derivative matrix $\hat{\ppvec{D}}$ is not DOF-local, since the contribution to point $(i,j,k)$ depends on the fluxes along the summation indices $\alpha, \beta, \gamma$ in \cref{eq:dgsem}, which are the linewise connections originally shown in \cref{fig:dg_cube}.
This introduces complications when trying to parallelize this routine on GPUs, because the update for a single interpolation point exhibits a wide stencil.
Our approach was to take advantage of the shared memory of the GPU, such that one thread block of the GPU gets assigned a single DG element.
This thread block then first computes all fluxes in the element, stores them in shared memory and then applies the derivative matrix to these fluxes stored in shared memory.
This ensures that the volume fluxes as intermediate results do not have to be moved to main memory, every flux has to be computed only once, and the full parallelism of the algorithm can be exploited.
If the split-flux formulation is employed, the algorithmic structure remains almost identical, but minor modifications are necessary, as already discussed in \cref{sec:dgsem}.

\subsubsection*{\code{ProlongToFace}}
The \code{ProlongToFace} routine interpolates the solution for $\ppvec{U}$, which are defined in the reference element, to the element faces.
The solution on the faces is denoted as $\tilde{\ppvec{U}}^{L/R}$, the solution from the left and right neighbor.
This is required later by the \code{FillFlux} routine to evaluate the fluxes across the faces.
For this, the polynomial basis from \cref{eq:ansatz} is evaluated at the faces with the current coefficients $\hat{\ppvec{U}}$.
Additional mappings then allow to infer to which global face number the specific local face of the element belongs in order to store the solution in the correct place.
This mapping from element to face information is required due to the unstructured nature of the mesh.
This mapping entails two different steps.
First, it has to account for the orientation of the face-based coordinates to the element-based coordinate system.
Second, the relative rotation of a face with its neighbor has to be considered.
These transformations are performed here to ensure that the face information is already correctly aligned for the computation of the face fluxes in \code{FillFlux}.

\subsubsection*{\code{FillFlux}}
This routine computes the common flux $\ppvec{f}^*=\ppvec{f}^*(\tilde{\ppvec{U}}^{L},\tilde{\ppvec{U}}^{R})$ across the element faces based on their left and right solution using a Riemann solver.
Since the correct mapping and orientation of the neighboring faces is already ensured by the \code{ProlongToFace} routine, the Riemann solver can be applied directly.
The routine is thus DOF-local and easily parallelizable for GPUs, since the Riemann solver can be applied independently for each interpolation point on the faces.
In addition, this routine multiplies the flux with the surface element $\hat{s}$, which entails the size of the face.

\subsubsection*{\code{SurfInt}}
This routine integrates the fluxes on the faces of the element and adds their contribution to the right-hand side of the DGSEM.
Again, this routine entails a surface to volume mapping, due to the same reasons as in the \code{ProlongToFace} routine.
Since this routine is discussed at length in \cref{sec:dgsem}, we refer to this section for further details.

\subsubsection*{\code{ApplyJac}}
The \code{ApplyJac} routine divides the right-hand side of \cref{eq:dgsem} with the Jacobian $\mathcal{J}_{ijk}$ of the mapping for each individual interpolation point.
This operation is thus DOF-local in the sense that the division can be performed independently for each interpolation point.
This operation exhibits a low arithmetic intensity, since it performs only 5 floating point multiplications per interpolation point, but 6 floats to be copied from memory (one entry for each conserved variable in $\ppvec{U}$ and the (pointwise scalar) Jacobian $\mathcal{J}$).
Hence, further optimization of the GPU implementation could fuse this operation for instance with the priorly computed \code{SurfInt} to increase the number of operations performed for each byte of data transferred.

\bibliographystyle{elsarticle-num-names}
\bibliography{bibliography}

\end{document}